\documentclass[10pt, a4paper,twocolumn]{article}
\usepackage{times}
\usepackage[T1]{fontenc}
\usepackage[utf8]{inputenc}
\usepackage{hyperref}
\hypersetup{colorlinks=true,citecolor=blue,linkcolor=blue,urlcolor=blue}
\usepackage[top=1in, bottom=1.25in, left=.75in, right=.75in]{geometry}
\usepackage{natbib} % [colon][super] Harcard bibtex
\usepackage{verbatim}
\usepackage{amssymb}
\usepackage{amsthm}
\usepackage{float}
\usepackage{enumerate}
\usepackage[titletoc]{appendix}
\usepackage{graphicx} %For figures
\usepackage{amsmath} %For matrix displays
\usepackage{nomencl}

\usepackage{relsize}
\usepackage{units} % set units and nicefrac (nicefrac package included)
\usepackage{bm} %bold math
\usepackage{setspace}
\usepackage[font=normal]{caption} %[skip = 2p]
\usepackage[belowskip=10pt,aboveskip=1pt,subrefformat=parens]{subcaption} %[skip = 2p]

\usepackage{xfrac}%NB uses a lot of compilation time
\usepackage{stmaryrd} %brackets
\usepackage{xcolor} %used in eg. pdfcomment
\numberwithin{equation}{section} %sets equation numbers <chapter>.<section>.<index>
\numberwithin{figure}{section} %sets figure numbers <chapter>.<section>.<index>

\newcommand{\pp}{\partial}
\newcommand{\dd}{\mr d}
\newcommand{\pdiff}[2]{\frac{\partial #1}{\pp #2}}
\newcommand{\ddiff}[2]{\frac{\mathrm d #1}{\dd #2}}

\newcommand{\brac}[1]{{ \left(  #1 \right) }}
\newcommand{\sqbrac}[1]{\left[ #1 \right]}

\newcommand{\ie}{\textit{i.e.}}
\newcommand{\egg}{\textit{e.g.}}

\let\underscore\_
\renewcommand{\_}[1]{_\mathrm{#1} }

\let\texthat\^
\renewcommand{\^}[1]{^\mathrm{#1} }

\newcommand{\of}[1]{\mbox {\smaller$\brac{#1}$}}

\newcommand{\wt}{\tilde}
\newcommand{\wh}{\hat}
\newcommand{\ol}{\overline}
\newcommand{\mr}{\mathrm}
\newcommand{\mc}{\mathcal}
\renewcommand{\l}{\_{\ell}}
\newcommand{\g}{\_g}
\renewcommand{\k}{_\phaseindex}

\renewcommand{\a}{A}

\newcommand{\dx}{{\Delta x}}
\newcommand{\dt}{{\Delta t}}
\newcommand{\ba}{\begin{align}}
\newcommand{\ea}{\end{align}}
\newcommand{\bseq}{\begin{subequations}}
\newcommand{\eseq}{\end{subequations}}

\newcommand{\jm}{_{j-1}}
\newcommand{\jp}{_{j+1}}
\newcommand{\jmh}{_{j-\frac12}}
\newcommand{\jph}{_{j+\frac12}}
\newcommand{\jpmh}{_{j\pm\frac12}}
\newcommand{\kj}{_{\phaseindex,j}}

\newcommand{\kjm}{_{\phaseindex,j-1}}
\newcommand{\kJ}{_{\phaseindex,J}}
\newcommand{\kJm}{_{\phaseindex,J-1}}

\newcommand{\lj}{_{\ell,j}}
\newcommand{\gj}{_{\mr g,j}}

\newcommand{\mcS}{\mc S}

\newcommand{\h}{\frac12}

\newcommand{\n}{^n}
\newcommand{\nn}{^{n+1}}
\newcommand{\nm}{^{n-1}}

\newcommand{\diff}[1]{\sqbrac{#1}}

\newcommand{\diffk}[1]{\diff{#1}\g\^\ell}
\renewcommand{\a}{a}
\newcommand{\al}{\a\l}
\newcommand{\ag}{\a\g}
\newcommand{\ak}{\a\k}
\newcommand{\area}{\mc A}

\newcommand{\const}{\mr{const}}
\newcommand{\Em}{\mbox{\sc{e}-}}
\newcommand{\Ep}{\mbox{\sc{e}}}
\newcommand{\G}{\mathbb G}
\newcommand{\C}{\mathbb C}

\newcommand{\alj}{a_{\ell,j}}

\newcommand{\aljp}{a_{\ell,j+1}}
\newcommand{\ulj}{u_{\ell,j}}

\newcommand{\A}{A}
\newcommand{\Al}{\A\l}
\newcommand{\Ag}{\A\g}

\newcommand{\br}{\brac}
\newcommand{\brt}[1]{{\brac{#1}\!\!}}
\newcommand{\bw}{\bm w}
\newcommand{\bv}{\bm v}
\newcommand{\bW}{\bm W}
\newcommand{\bV}{\bm V}
\newcommand{\bvt}{\wt \bv}
\newcommand{\bvh}{\wh \bv}
\newcommand{\bwt}{\wt \bw}
\newcommand{\bwh}{\wh \bw}

\renewcommand{\bf}{\bm f}
\newcommand{\bF}{\bm F}
\newcommand{\bFr}{\bm F\_r}
\newcommand{\bft}{\tilde \bf}
\newcommand{\bs}{\bm s}
\newcommand{\bS}{\bm S}

\newcommand{\m}{^*}
\newcommand{\s}{s}
\renewcommand{\S}{S}
\newcommand{\bE}{\bm E}
\newcommand{\bEdisc}{\bm E\^d}
\newcommand{\dS}{\S'}
\newcommand{\Jr}{\J\_r}
\newcommand{\dJr}{\Jr'}
\newcommand{\dJrd }{\J\_{r,d}'}
\newcommand{\dJ}{\J'}
\newcommand{\dSd}{S\_d'}

\newcommand{\SQl}{\mc \S_{\Q\l}}
\newcommand{\SQg}{\mc \S_{\Q\g}}

\newcommand{\SAl}{\mc \S_{\A\l}}

\newcommand{\Q}{Q}
\newcommand{\Ql}{\Q\l}
\newcommand{\Qg}{\Q\g}
\newcommand{\Qm}{\mc Q}
\newcommand{\Qkr}{\Q_{\phaseindex,\mr r}}
\newcommand{\q}{q}
\newcommand{\ql}{\q\l}
\newcommand{\qg}{\q\g}
\newcommand{\qk}{\q\k}
\renewcommand{\j}{j}
\newcommand{\J}{J}
\newcommand{\cnu}{c_\nu}
\newcommand{\cnudisc}{\cnu\^d}
\newcommand{\ddisc}{\delta\^d}

\newcommand{\cdisc}{c\^d}

\newcommand{\my}{m_y}
\newcommand{\mx}{m_x}
\newcommand{\HofAl}{\mc H}
\newcommand{\dHdAl}{\HofAl'}
\newcommand{\Jacr}{\pdiff \bFr \bV}  %\bVr
\newcommand{\Jac}{\pdiff \bF \bV}
\newcommand{\jac}{\pdiff \bf \bv}
\newcommand{\dSdV}{\pdiff \bS \bV }
\newcommand{\I}{\mathbb I}
\newcommand{\phix}{\phi_x}
\newcommand{\phit}{\phi_t}
\newcommand{\p}{p}
\newcommand{\ii}{\rho\q}
\newcommand{\mm}{\rho\a}

\newcommand{\Uk}{U\k}
\newcommand{\Ukr}{U_{\phaseindex,\mr r}}
\newcommand{\Ur}{U\_r}

\newcommand{\LL}{\mathbb L}
\newcommand{\inv}{^\text{-1}} 
\newcommand{\PP}{\mathbb P}
\newcommand{\Lamb}{\Lambda}

\newcommand{\CFL}{\text{CFL}}
\newcommand{\USL}{\Ql/\area}
\newcommand{\USG}{\Qg/\area}

\newcommand{\phaseindex}{\kappa}
\newcommand{\symkappa}{\varkappa}
\newcommand{\imunit}{\mr i}
\newcommand{\eEuler}{\mr e}

\newcommand{\markercolor}{black}

\newcommand{\trykkext}{pdf}

\theoremstyle{plain}

\newtheorem{example}{Example}

\newtheorem{remark}{Remark}

\numberwithin{equation}{section}

\title{The Kelvin-Helmholtz/von Neumann Stability of Discrete Representations of the Two-Fluid Model for Stratified Two-Phase Flow}% 

\author{
\textsc{A.H.~Akselsen}\\[1ex]
\normalsize Norwegian University of Science and Technology, \\
\normalsize Department of Energy and Process Engineering, Kolbj{\o}rn Hejes v.\ 1B, 7491 Trondheim, Norway. \\ 
\normalsize \href{mailto:andreas.h.akselsen@ntnu.no}{andreas.h.akselsen@ntnu.no}
}

\date{ }

\begin{document}

\maketitle

\begin{abstract}
Many dynamic pipe flow simulator tools are capable of predicting the onset of hydrodynamic flow instability through detailed simulation. 
These instabilities provide a natural mechanism for flow regime transition. 
The quality and reliability of flow predictions are however strongly dependent upon the numerics within these simulator tools, the scheme type and resolution in particular. 
%Many modern simulation tools to date still rely on low-precision schemes like the upwind and Lax-Friedrich schemes for capturing flow instabilities on discrete meshes. 

A Kelvin-Helmholtz stability analysis for the differential two-fluid model is in the present work presented and extended to discrete representations of said model. 
%Linear stability analyses are performed on the discretized model representations and presented in the present work. 
This analysis provides algebraic expressions which give instantaneous, quantitative information into i) when a studied scheme will predict linear wave growth, ii) the rate of growth and the expected growing wavelength, and iii) 
%the phase velocity of disturbance waves. 
the wave speeds.
These stability expressions adhere to a wider family of finite volume methods, directly applicable to any specific formulation within this group.
Both the spatial and temporal discretization are found to play decisive roles in a method's predictive capability.
Fundamental aspects of how numerical errors from the temporal integration affects the predicted stability are explored. 
Numerical errors are observed to manifest in increased, as well as reduced, wave growth. 
Low-frequency growth from numerical errors is not always easily distinguished from physical wave growth.
The linear analysis is demonstrated to be useful in understanding the predictions made by simulator tools, and in choosing the appropriate numerical method and simulation parameters for optimizing the simulation efficiency and reliability.
%The analyses also provide insight into the mechanism of hydrodynamic wave instability.
%
%the influence of the time integration.
% and into the differences between viscous and inviscid flows in this respect. 
%For example, it is found that numerical viscosity can promote premature instability in otherwise low-viscosity flows. 
%For example, numerical viscosity, most commonly regarded to have a diffusive and stabilizing influence, is found 
%Results are compared with actual simulations. These analyses are of interest in understanding the predictions made by simulator tools and in choosing the appropriate methods, grid resolution and time steps with respect to prediction reliability and simulation efficiency.

\end{abstract}

%\tableofcontents
%
%\input{nomenclature}
%{\small
%\printnomenclature%[0.2 cm]
%}

%\linenumbers
%\nolinenumbers

\section{Introduction}

%The two-fluid model is well-known for having complex eigenvalues if the momentum exchange between the two phases is insufficient \cite{Gidaspow_fluidization}. Models with complex eigenvalues are not hyperbolic and cannot constitute a well-posed initial value problem.
Stability analyses of the two-fluid model have been performed by numerous authors.
For example, 
\textcolor{\markercolor}{
Taitel and Dukler~\cite{Taitel_Dukler_friction_factor}
used linear theory with a simplified, inviscid two-fluid model to predict flow regime transition to slug flow.
}%
Barnea and Taitel~\cite{Barnea_VKH_and_IKH} presented a derivation of the Kelvin-Helmholtz stability criterion for viscous flows (henceforth abbreviated the VKH criterion) and examined the non-linear flow development through simulation \cite{Barnea_simplified_with_characteristics}. 
Barnea also performed a stability analysis on a discrete upwind type scheme of a simplified version of the two-fluid model for annular flow \cite{Barnea_Discrete_annular_instability}. 
Here it was shown again how an intrinsically unstable, ill-posed differential model may display stable behaviour if provided with sufficient numerical diffusion. (The annular interface is inherently unstable locally though it may be stable in a statistical sense.)
Barnea argued  that the discrete model can be regarded as a legitimate model for the average flow, even though the differential model  is ill-posed.
%\\

Issa and Kempf~\cite{Issa_two_phase_capturing} have been credited with first demonstrating that the predicted wave growth from transient simulations of the full two-fluid model coincides with the wave growth from Kelvin-Helmholtz theory, and suggested that such simulated wave growth gives a natural transition into a wavy or slugging flow regime.

In ~\cite{Stewart_stability_well_illposed}, Stewart presented a von Neumann analysis on two variants of
the two-fluid model, one with a term exchanging momentum between the phases and one without.
The two-fluid model is known for obtaining complex eigenvalues if the momentum exchange is insufficient, which means that the model can no longer be deemed part of a well-posed hyperbolic initial value problem \cite{Gidaspow_fluidization}.
%both well-posed and ill-posed two-fluid models. 
Stewart showed that \textit{well-behaved} discrete solutions, \ie, steady flow solutions, were obtainable on discretizations of non-hyperbolic systems, provided the mesh resolution was restricted. 
%A donor-cell, or upwind, type discretization was here used with a staggered grid.

Liao et al.\ \cite{Liao_von_Neumann} performed a linear stability study on a discrete two-fluid model with a staggered grid arrangement, comparing various interpolations for the convection term.
This analysis was limited to implicit time integration, considering  the convection terms only.
Numerical errors arising from the dislocation of staggered information, as well as form the conservative formulation, appears to have been neglected.
% in order to determine which representation of the convection term would provide the most accurate wave growth predictions.
%
%Four discretizations were tested: the central difference discretization, the `first order upwind' or `donor-cell' discretization, second order upwind and the QUICK discretization. 
The paper concluded that the central difference discretization was superior to the first and higher order non-centred interpolations.
%The paper concluded with the central difference discretization being superior in this respect,
%though this analysis appears to be limited to a specific staggered stencil with implicit time integration, considering  different interpolations of the convection terms only.
%Two-fluid model representations both with and without friction was considered.
%Special emphasis was placed on the difference in discrete response of the two-fluid model with source terms (friction) as opposed to without. 
Liao et al.\ also examined the evolution of the wavelength distributions from a random initial disturbance,
and the behaviour as the model turns ill-posed.
%and the transition to ill-posed flow.

\textcolor{\markercolor}{
The light water reactor safety analysis codes RELAP5 and CATHARE have also been studied with von Neumann analysis by Pokharna et al..\ \cite{Pokharna_von_Neumann_RELAP5_CATHARE}, looking into how numerical diffusion and terms added to achieve model hyperbolicity affect stability predictions.
They found the numerical regularization to be dominant in the cases studied.
Fullmer et al.~\cite{Fullmer_VKH_von_Neumann} preformed similar analyses on an upwind discretization of the two-fluid model, demonstrating the mesh size dependency of the predicted stability in both the linear and nonlinear range.
A Reynolds stress modelling was shown provide grid independent regularization.
}%
\\

%  The study did not address High-wavenumber numerical issues associated with central difference convection, such as the Gibbs oscillation phenomena associated with discontinuities.
%\\

%The doctoral thesis of Renault~\cite{LASSI} contained a stability analysis on a particular upwind type, staggered grid discretization of the two-fluid model. The spatial numerical diffusion and dispersion effects were included into a extended dispersion equation to yield a stability criterion analogous to the VKH criterion for this upwind scheme. The resulting criterion was valid for low wavenumbers only.

The present article focuses on providing general theory for a wider family of discrete two-fluid model representations. %, which will prove useful in assessing a method's stability and predictive capability.
This will be done by relating the predicted growth and decay of the discrete representations directly to the growth results of the Kelvin-Helmholtz analysis of the differential two-fluid model. 
Linear theory of the type here presented is demonstrated to be powerful tool when it comes to assessing the predictive capability of any chosen discrete representation, providing support with decisions related to the parametric setup prior to simulation and interpreting the simulation results.
\textit{Predictive capability} here refers to the reliability and accuracy with which a discrete representation predicts wave growth or decay under limited computational resolution. 
It will be shown that the growth and dispersion response of discrete representations is perfectly analogous to that of the differential model. 
What's more, the differential Kelvin-Helmholtz expression directly provides that growth and dispersion which will be predicted by the discrete methods in the linear range, provided these representations uses the same discrete differentiations all over.

\section{The Two-Fluid Model}
\label{sec:base_model}

\begin{figure}[h!ptb]%
\centering
%\includegraphics[width=.35\columnwidth]{./images/pipe_cross_section3.pdf}%
%\hfill
%\includegraphics[width=.6\columnwidth]{./images/pipe_sideways.pdf}%
\begin{minipage}[c]{.35\columnwidth}
\includegraphics[width=\columnwidth]{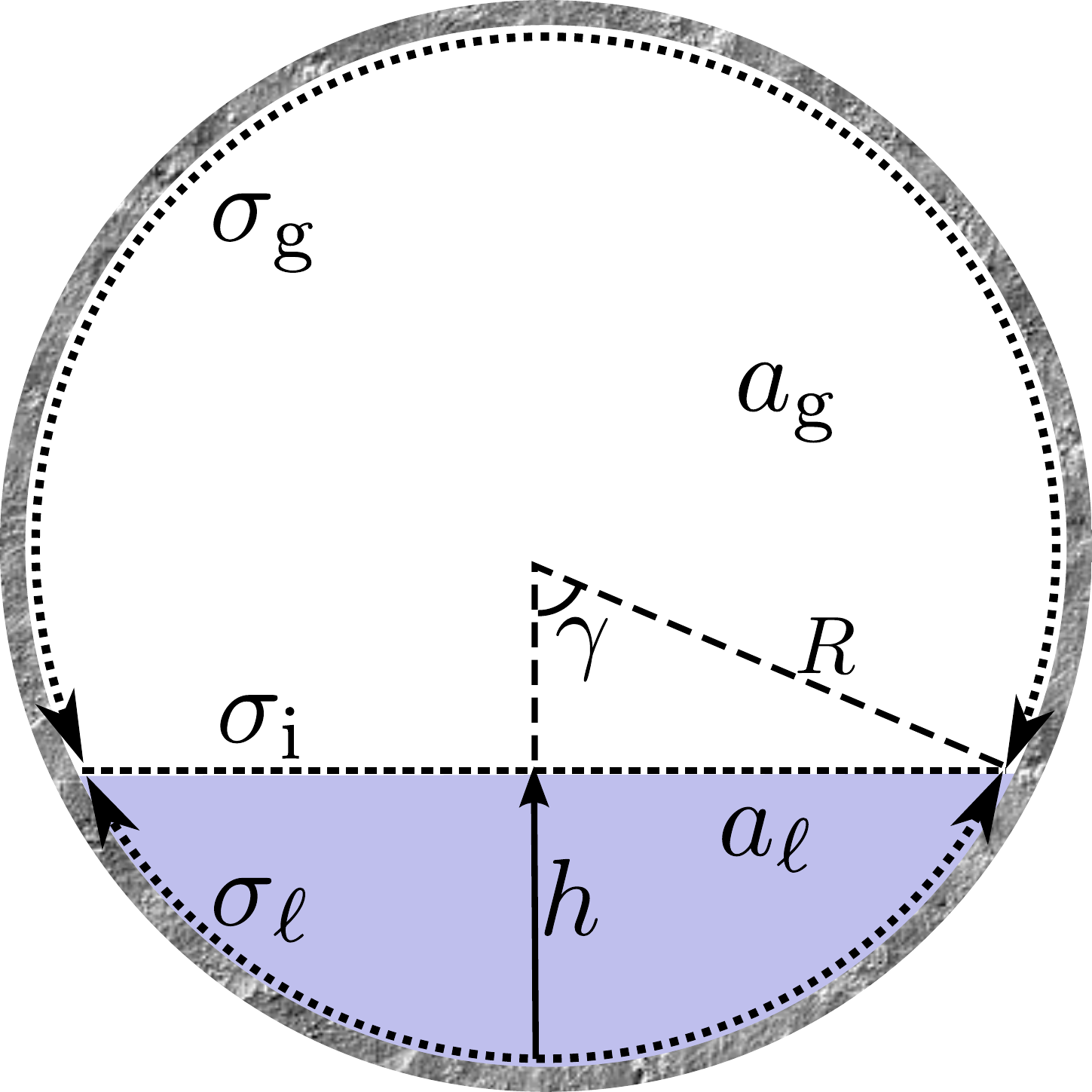}%
\end{minipage}
\hfill
\begin{minipage}[c]{.6\columnwidth}
\includegraphics[width=\columnwidth]{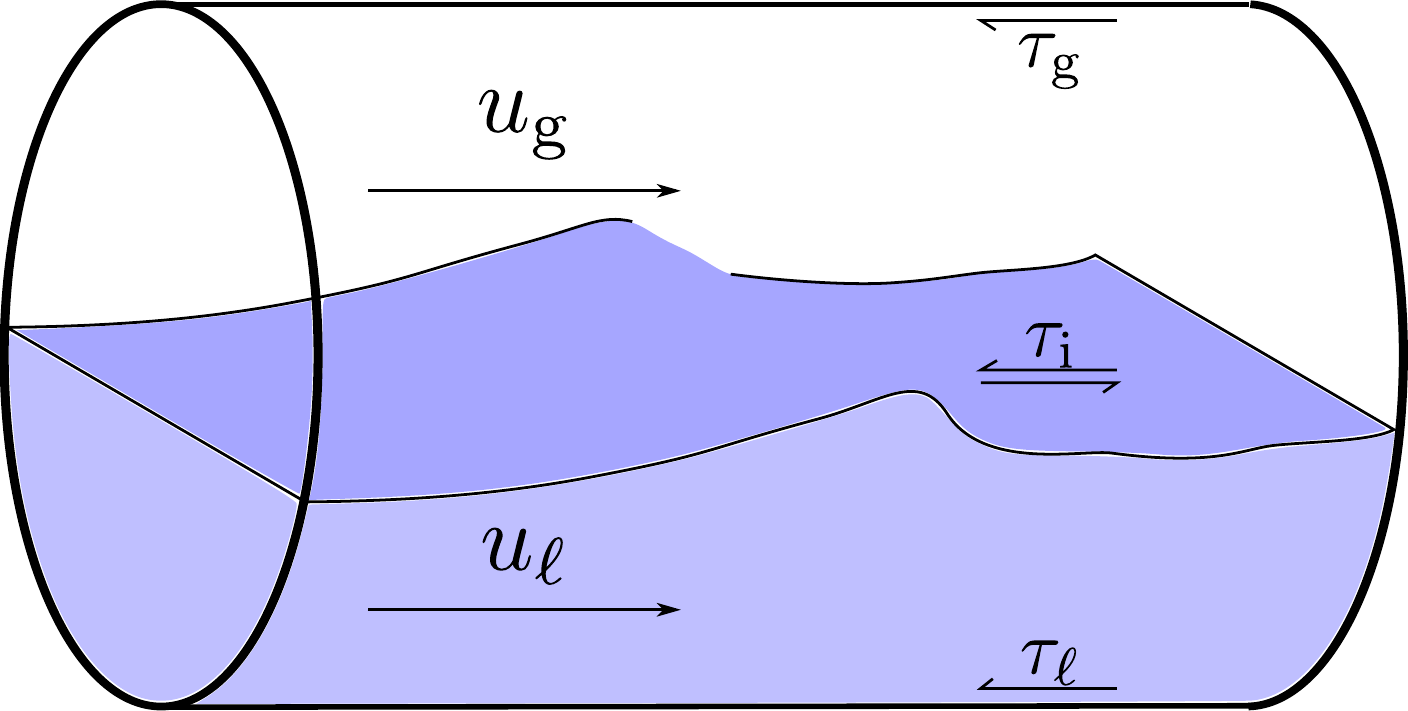}%
\end{minipage}
\caption{Pipe cross-section}%
\label{fig:cross_section}%
\end{figure}

The compressible, adiabatic, equal pressure four-equation two-fluid model for stratified pipe flow results from an averaging of the conservation equations across the cross-section area.  %Written with generic phase indices $\phaseindex = \{\ell, \mr g\}$ for liquid and gas, 
The model is commonly written
\begin{subequations}
\begin{align}
%	&\br{\rho\k  a\k }_t+  \brac{\rho\k  a\k  u\k}_x = 0, \label{eq:base_model_with_pressure:mass} \\
%	& \br{\rho\k   a\k  u\k}_t + \brac{\rho\k  a\k  u\k^2}_x +  a\k p_{\mr i, x} +g_y  \rho\k   a\k  h_x  = s\k, \\ \label{eq:base_model_with_pressure:mom} 
%	&\pp_t \rho\k  \ak  + \pp_x\rho\k  a\k  u\k = 0,  \label{eq:base_model_with_pressure:mass} \\
%	&\pp_t \rho\k   \ak  u\k + \pp_x \rho\k  a\k  u\k^2 + \ak \pp_x p\_i + \ak \rho\k g \cos\theta\, \pp_x h = s\k, \\ \label{eq:base_model_with_pressure:mom}
	&\pp_t \br{\rho \a  }\k + \pp_x \br{\rho \a u}\k = 0,  \label{eq:base_model_with_pressure:mass} \\
	&\pp_t \br{\rho \a u}\k + \pp_x \br{\rho \a u^2}\k + \ak\pp_x p\_i + \rho\k  \ak g \cos\theta\, \pp_x h = s\k, \\ \label{eq:base_model_with_pressure:mom}
%	&\pp_t \br{\rho \a  }\k + \pp_x \br{\rho \q}\k = 0,  \label{eq:base_model_with_pressure:mass} \\
%	&\pp_t \br{\rho \q}\k + \pp_x \br{\rho \q u}\k + a\k \pp_x p\_i + \rho\k  \ak g \cos\theta\, \pp_x h = s\k, \\ \label{eq:base_model_with_pressure:mom}
	&\a\l+\a\g=\area,\\
	&p\_i = \mc P\g\of{\rho\g} = \mc P\l\of{\rho\l}.
\end{align}%
\label{eq:base_model_with_pressure}%
\end{subequations}%
Field $\phaseindex$, occupied by either gas, $\phaseindex = \mr g$, or liquid, $\phaseindex = \ell$, is segregated from the other field. Subscript $\mr i$ indicates the fluid interface; see Figure~\ref{fig:cross_section}.
%Equations~\eqref{eq:base_model_with_pressure:mass} and \eqref{eq:base_model_with_pressure:mom} hold for both the gas field, $\phaseindex = \mr g$, and the liquid field, $\phaseindex = \ell$.
$p\_i$ is here the pressure at the interface, assumed the same for each phase and given by some equation of state $\mc P$.  %Surface tension is here neglected.
$h$ is the height of the interface from the pipe floor, and the term in which it appears originates from approximating a hydrostatic wall-normal pressure distribution. 
%$\ak$ is the cross-section area within field $\phaseindex$, and $u\k$ and $\rho\k$ are the mean fluid velocity and density in these fields. 
$u\k$ and $\rho\k$ are the mean fluid velocity and density in field $\phaseindex$. 
%See Figure~\ref{fig:cross_section} for illustration.
The momentum sources are 
$s\k = -\tau\k \sigma\k \pm \tau\_i \sigma\_i - \a\k \rho\k g \sin\theta$,
where $\tau$ is the skin frictions at the walls and interface. $\theta$ is the pipe inclination, positive above datum, and $g$ is the gravitational acceleration.
%Internal mass sources $s_k\^{mass}$ are commonly zero.
\\

%
%Figure~\ref{fig:cross_section} illustrates the pipe geometry and some of the quantities appearing the two-fluid model.
%
%\\

%Field $\phaseindex$, occupied by either gas $\phaseindex = \mr g$ or liquid $\phaseindex = \ell$, is segregated from the other field. Subscript $\mr i$ indicates the fluid interface.
%The circular pipe geometry itself enters into the modelling through the relation between the level height $h$, the specific areas $\ak$ and the perimeter lengths $\sigma\k$ and $\sigma\_i$.
%These are algebraically interchangeable through a geometric function
%\\

The circular pipe geometry itself enters into the modelling through the relation between the level height $h$, the specific areas $\ak$ and the peripheral lengths $\sigma\k$ and $\sigma\_i$.
These are algebraically interchangeable through a geometric function
%\\
%
\begin{equation}
h = \HofAl\of \al
\label{eq:Alofh}
\end{equation}
whose derivative is  $\dHdAl = 1/\sigma\_i$.
See \egg \cite{Akselsen_char_TEMP} for expressions of the geometric relationships.
\\

%%
%\begin{equation}
%h = \HofAl\of \al.
%\label{eq:Alofh}
%\end{equation}
%%
%Only the inverse of $\HofAl$ is an explicit expression
%%
%\begin{align*}
%\HofAl\inv\of h &= R^2\br{\gamma - \nicefrac12 \sin 2\gamma},
%&
%\gamma\of h &= \arccos\br{1-\frac h R},
%\end{align*}
%%
%but Biberg's approximation \cite{Biberg_h_of_al}
%%
%\begin{align*}
%\HofAl\of \al  &= R \br{1-\cos\gamma},
%\\
%\gamma & \approx\ \pi \alpha\l + \br{\frac{3\pi}{2}}^{\!\!\nicefrac13}\!\br{1-2\alpha\l+\alpha\l^{\nicefrac13}-\alpha\g^{\nicefrac13}}\\
%&\quad\;\; -0.005\,\alpha\l\alpha\g\br{\alpha\g-\alpha\l}\br{1+4\br{\alpha\l^2+\alpha\g^2}^2},
%%\\
%\end{align*}
%is very accurate.
%$R$ is here the pipe inner radius, $\gamma$ the interface half-angle and $\alpha\k = \ak/\area$ is the phase fraction.
%The function derivative is $\dHdAl = 1/\sigma\_i$
%and the perimeter lengths are 
%\begin{align*}
%\sigma\l &= 2 R \gamma,
%&
%\sigma\g &= 2R\br{\pi-\gamma},
%&
%\sigma\_i &= 2R\sin \gamma.
%\end{align*}
%\\

Fluid compressibility is commonly ignored when considering the surface wave stability of \eqref{eq:base_model_with_pressure}. 
This enables us to base the stability analysis on the incompressible two-fluid model, which has lower rank and a conservative form.
%Which model is actually being simulated, be it compressible or not, will only affect the predicted stability in as far as these choices affect the discretization strategies. This will be demonstrated later. 

%Now assuming both phases to be incompressible, we reduce the momentum equations with their respective mass equations.
%The pressure term is eliminated by dividing each momentum equation by their respective phase area and subtracting one from the other, resulting in
Assuming incompressible phases, the two-equation model is obtained by reducing the momentum equations with their respective mass equations and eliminating the pressure term between them, resulting in
\begin{equation}
\partial_t \bv + \partial_x \bf = \nu \pp_{xx}\bv + \bs
\label{eq:base_model}
\end{equation}
with conserved variables and fluxes
\begin{align*}
%\bv &= \big( \al, \diffk{\rho u}  \big)^T,
\bv &= \big( \al, \rho\l u\l-\rho\g u\g  \big)^T,
&
\bf &= ( \ql, \j )^T,
&
\bs & = (0,s)^T.
\end{align*}
Symbols for the flux and source components have here been defined and are
\begin{equation*}
\begin{gathered}
\qk = \ak u\k,
\qquad
%\j =  \diffk{\rho  {u^2/2}} +  \my h,
%\j =  \rho\l{ u\l^2}/{2} - \rho\g{ u\g^2}/{2} +  \my h,
\j =  \tfrac12\br{\rho\l u\l^2-\rho\g u\g^2 }+  \my h,
\\
%\s = -\mx -\diffk{\tau \sigma/a} + \tau\_i \sigma\_i \br{1/\al+1/\ag},
%\\
\s = -\mx - \frac{\tau\l \sigma\l}\al + \frac{\tau\g \sigma\g}\ag + \tau\_i \sigma\_i \br{\frac1\al+\frac1\ag},
%&&
\\
\nu = 0.
\end{gathered}
\end{equation*}
A dummy viscous term has been added to the system, the purpose of which lies in evaluating the artificial numerical viscosity present in some discrete representations.
%The shorthand
%\[
%	\diffk \cdot = (\cdot)\l-(\cdot)\g
%\]
%is useful throughout.
Specific weight coefficients have been grouped into $\mx = (\rho\l-\rho\g) g \sin \theta$ and $\my = (\rho\l-\rho\g) g \cos \theta$.
%$\sigma$ and $\tau$ respectively are the periphery lengths and skin frictions at the walls and interface. 
%Although really derivatives therefrom, these equations are here simply referred to as base mass and momentum equations.
%
The identities
\begin{align}
%\sumk (\ak,\ak u\k) =(\area,\Qm),
\al+\ag &= \area\of x, & \ql+\qg &= \Qm\of t,
\label{eq:sum_a_au}
\end{align}
%where the sum indicates summation over the two phases $k=\ell,\mr g$,
where the latter has been obtained from summing the two mass equations, close the model.
Both $\area$ and $\Qm$ are parametric.
% -- the former may be made to vary in space according to the geometry and the latter in time according to the mixture rate imposed upon the system. 
\\

Finally, the eigenstructure of \eqref{eq:base_model} is useful to know.
The Jacobian of $\bf$ is
\begin{equation}
	\pdiff \bf\bv = \frac{1}{\rho\m}
	\begin{pmatrix}
	\br{\rho u}\m & 1\\
	\symkappa^2 & \br{\rho u}\m 
	\end{pmatrix},
\label{eq:base_model:Jac}
\end{equation}
whose eigenvalues are
\begin{equation}
\lambda^\pm = \frac{(\rho u)\m \pm \symkappa}{\rho\m}.
\label{eq:eigenvalues}
\end{equation}
A new variable 
\begin{equation}
	\symkappa = \sqrt{\rho\m\my\dHdAl - \frac{\rho\l \rho\g}{\al \ag} \br{u\g-u\l}^2}.
	\label{eq:kappa}
\end{equation} %
has here been introduced along with the operator
\begin{equation}
\psi\m = \frac{\psi\l}{\al} + \frac{\psi\g}{\ag}.
\label{eq:asterisk_operator}
\end{equation}

\section{Kelvin-Helmholtz Stability }
\label{sec:VKH}
The viscous Kelvin-Helmholtz (VKH) stability analysis is here presented in some detail, 
which will later be related directly to the 
%This will prove beneficial later in the context of studying the 
stability of discrete representations.
%\remark{
%\label{rem:variable_change}
%The linearization step allows us to freely change disturbance variables through Jacobians evaluated steady state (constant).
%}
\\

\textit{Variable of the steady state $\bv=\bV$ will in the following be assigned upper-case symbols.}%
\footnote{\color{\markercolor}
The state $\bV$ could also be non-uniform provided the perturbation wavelengths are much smaller than the length scales of the flow state \cite{Pokharna_von_Neumann_RELAP5_CATHARE}.
}
The steady state solution $\bV$ satisfies the so-called \textit{holdup equation} 
%The steady state solution $\bV$ of \eqref{eq:base_model}, assigned upper-case symbols, satisfies the so-called \textit{holdup equation} 
%
\begin{equation}
\s\of\bV = \S = 0.
\label{eq:holdup_eq}
\end{equation}
linearizeing \eqref{eq:base_model} about the steady state, 
\[
%\bv = \bV + \bvt
\bv = \bV + \bvt,
\]
 yields
\begin{equation}
%\partial_t \bvt + \Jac \partial_x \bvt =  \pdiff \bS \bV \bvt.
\br{\I(\partial_t- \nu \pp_{xx}) + \Jac \partial_x - \dSdV}\bvt = \bm 0.
\label{eq:base_model_linearized}
\end{equation}
%
%Quite generally, a solution of system \eqref{eq:base_model_linearized} may be written
Let's briefly look at the general solution of \eqref{eq:base_model_linearized}. It may be written
\begin{equation}
%\bvt\of{x,t} = \sum_k \LL\inv \mr{disg}\br{\eEuler^{\lambda t}} \LL \bvh_{0,k} \eEuler^{\imunit kx}
%\bvt\of{x,t} = \sum_k \PP\inv \mr{diag}_i\br{\eEuler^{\lambda_{i,k} t}} \PP \,\bvh_{0,k} \eEuler^{\imunit kx},
%\bvt\of{x,t} = \sum_k \PP\inv \eEuler^{\Lamb_{k} t} \PP \,\bvh_{k}^0 \eEuler^{\imunit kx}
\bvt\of{x,t} = \sum_k \PP\inv \eEuler^{\imunit k(x-\C_{k} t)} \PP \,\bvh_{k}^0
\label{eq:general_sol_lin_syst}
\end{equation}
%
%where the eigenvalues $\lambda$ and the eigenvectors in $\PP$ 
where 
$
\eEuler^{-\imunit k\C_k t} = \mr{diag}_\p(\eEuler^{-\imunit k c_{k,\p} t}),
$
$c_{k,\p}$ being the $\p$-th eigenvalues of 
$
\mathbb H_k =  \Jac - \imunit k\nu\I + \frac \imunit k \pdiff \bS \bV
$
and $\PP$ containing their corresponding eigenvectors. 
%%the respective eigenvalues and eigenvectors in $\C$ (diagonal) and $\PP$ 
$\bvh_{k}^0$ are the Fourier modes of the initial conditions.
\begin{proof}
Solution \eqref{eq:general_sol_lin_syst} satisfies the initial conditions by virtue of $\bvh_{k}^0$ being the Fourier modes of these and $\eEuler^{-\imunit k\C \cdot 0} = \I$.
Further we have
\begin{align*}
\pp_t \PP\inv  \eEuler^{\mathbb -\imunit k\C t} \PP &= \PP\inv \mr{diag}_\p(-\imunit k c_\p \eEuler^{-\imunit k c_\p t}) \PP \\
&=\PP\inv (-\imunit k \C) \eEuler^{\mathbb -\imunit k\C t} \PP  = -\imunit k \mathbb H \PP\inv  \eEuler^{\mathbb -\imunit k\C t} \PP.
\end{align*}
Directly inserting \eqref{eq:general_sol_lin_syst} into \eqref{eq:base_model_linearized} then yields
\[
\sum_k\! \br{\!-\imunit k\mathbb H_k +k^2\nu \I  +\Jac \imunit  k - \pdiff \bS\bV} \PP\inv\! \eEuler^{\imunit k(x-\C_{k} t)} \PP\,\bvh_k = \bm 0.
\]
The bracket term cancels at each wavenumber due to the definition of $\mathbb H_k$.
\end{proof}
So, the linear response of the system will be through the growth and dispersion of a number of linear waves.
We will not bother too much with this general solution, but are
interested in the stability behaviour of \eqref{eq:base_model_linearized} -- stable flow occurs if the real component of all eigenvalues is negative or zero.
The solution~\eqref{eq:general_sol_lin_syst} is just a linear combination of waves; we re-write it to the form 
\begin{equation}
%\bvt\of{x,t} = \sum_\p\bvh_{\p} \eEuler^{ik_\p(x-c_{\p}t)}. 
\bvt\of{x,t} = \sum_k \sum_{p=1,2} \bvh_{k,\p} \eEuler^{\imunit k(x-c_{k,\p}t)}.
%\bvt\of{x,t} = \sum_k \br{\bvh_{k,+} \eEuler^{\imunit k(x-c_{k,+}t)} + \bvh_{k,-} \eEuler^{\imunit k(x-c_{k,-}t)}}.
\label{eq:bvt}
\end{equation}
%
%where $c_\p = \frac \imunit k \lambda_\p$.
%Two celerities $c_\p$ here exist for every wavenumber $k$ because the rank of \eqref{eq:base_model_linearized} is two.
Inserting \eqref{eq:bvt} into \eqref{eq:base_model_linearized} yields the algebraic system
\begin{equation}
%\pdiff\bE \bV \bvt = 0
%\sum_\p\pdiff{\bE_\p} \bV \bvh_\p  \eEuler^{ik_\p(x-c_\p t)} = 0,
\sum_k \sum_{p=1,2} \pdiff{\bE_{k,\p}} \bV \bvh_{k,\p}  \eEuler^{\imunit k(x-c_{k,\p} t)} = \bm 0,
%\sum_k \br{ \pdiff{\bE_{k,+}} \bV \bvh_{k,+}  \eEuler^{\imunit k(x-c_{k,+} t)}+ \pdiff{\bE_{k,-}} \bV \bvh_{k,-}  \eEuler^{\imunit k(x-c_{k,-} t)}} = 0,
\label{eq:base_model_linearized_dEdV}
\end{equation}
with
\begin{equation}
%\bE = \bV(\delta_t - \nu \delta_{xx}) + \bF \delta_x - \bS .
%\bE_\p= \bV(\delta_t^\p - \nu \delta_{xx}^\p) + \bF \delta_x^\p - \bS .
\bE_{k,\p}= \bV(\delta_t^{k,\p} - \nu \delta_{xx}^{k}) + \bF \delta_x^{k} - \bS .
\label{eq:E}
\end{equation}
%
%For the purpose of comparing with the discrete analyses refrain from inserting the wave modes and have instead defined 
Each $\p$-term must solve  \eqref{eq:base_model_linearized_dEdV} individually if the sum is to be a solution at all times.
Suppressing both sum indices we simply write
\begin{equation}
\pdiff{\bE} \bV \bvh= \bm 0.
%\pdiff{\bE_\p} \bV \bvh_\p=0.
%\pdiff{\bE_{k,p}} \bV\bvh_{kp} =0
\label{eq:base_model_dEdV_nosum}
\end{equation}
%
%The $\p$-indexing is from here dropped. 
The $\delta$-operators appearing in \eqref{eq:E}, accounting for the effect of the partial derivatives, are defined
\begin{equation}
%\delta^{i,k} \equiv (\pp \wt \psi_{i,k})/{\wt\psi_{i,k}}.%
\delta \equiv \frac{\pp  \exp {\imunit k(x-c t)}}{ \exp{\imunit k(x-ct)}}.
%\delta \equiv \frac{\pp \,  \eEuler^{\imunit k(x-c t)}}{ \eEuler^{\imunit k(x-ct)}}.
%\delta \equiv \br{\pp \,  \eEuler^{\imunit k(x-c t)}}/ \eEuler^{\imunit k(x-ct)}.
\label{eq:delta}
\end{equation}%
Note that these are simple scalars effectively flipping the various terms straight angles in the complex plane:
\begin{align*}
\delta_t &= -\imunit kc,& \delta_x &= \imunit k, &\delta_{xx} &= -k^2.
\end{align*}%
Using $\delta$ operators will 
%prove insightful when comparing with the discrete analysis later. 
allow solutions to be extended directly to discrete representations.
Because  \eqref{eq:base_model_dEdV_nosum} is linear we may express it uniquely in terms of one of the disturbance properties, 
%say $\hat\a\l = \pdiff \Al\bV \bvh\neq 0$, 
say $\hat\a\l: \pdiff \bV \Al   \hat\a\l = \bvh$, 
yielding
\begin{equation}
%\ddiff \bE \Al = 0.
\bE' =\bm  0,
\label{eq:dE_eq_0}
\end{equation}
where $\Psi' \equiv \ddiff \Psi \Al$.
%
%%
%\begin{equation}
%\ddiff {}\Al \sqbrac{\bV \br{\partial_t- \nu \pp_{xx}} + \bF\pp_x  - \bS}\wt \a\l = 0,%\qquad \wh \a\l = \pdiff \Al \bV \bvh \neq 0.
%%\ddiff {\br{\partial_t- \nu \pp_{xx}}\bV + \bF  - \bS}\Al  \wt \a\l = 0,\qquad \wh \a\l = \pdiff \Al \bV \bvh \neq 0.
%\label{eq:base_model_linearized_dAl}
%\end{equation}
%%
%where $\wh \a\l = \pdiff \Al \bV \bvh \neq 0$.
%The deferential operators effect on $\alt$ is to flip it through multiplication with complex constants.
%For the purpose of later comparisons with the discrete analysis we refrain from inserting the wave mode and define instead
%%These constants may be represented by defining
%%
%\begin{equation}
%\delta \equiv \frac{\pp \alt}{\alt}.
%\label{eq:}
%\end{equation}
%%
%%which is always well defined.
We further define a viscous phase celerity
\begin{equation}
\cnu \equiv - \frac{\delta_t -  \nu \delta_{xx}}{\delta_x}, 
%\frac{-\pp_t \alt + \nu \pp_{xx}\alt}{\pp_x\alt} = c + \imunit  k \nu
\label{eq:cnu}
\end{equation}
which evaluates to  $\cnu =  c + \imunit  k \nu$.
%Equation \eqref{eq:dE_eq_0} is now written
Inserting \eqref{eq:E}  and \eqref{eq:cnu} into \eqref{eq:dE_eq_0} now yields
\begin{equation}
%\br{\bFr'\pp_x + \bS'}\alt = 0, 
\bFr'\delta_x - \bS' = \bm 0. 
\label{eq:disp_eq_ddAl}
\end{equation}
The components of $\bFr$ are the fluxes in a relative frame, moving with (complex) velocity $\cnu$.  
$\bFr$ equals $\bF$ with the relative velocities
\[
\Ukr = \Uk-\cnu
\]
replacing $\Uk$.
Since the mass equation contains no source term,  the first component of \eqref{eq:disp_eq_ddAl}, combined with \eqref{eq:sum_a_au},  yields directly 
\begin{equation}
\Qkr = \A\k\Ukr = \const., %\text{(complex)}%\in \mathbb C,
\label{eq:Qr_const}
\end{equation}
which relates both velocity components to $\Al$.
The second component of \eqref{eq:disp_eq_ddAl} yields the \textit{dispersion equation}
\begin{equation}
\dJr\, \delta_x  - \dS = 0,
\label{eq:disp_eq}
\end{equation}
where $\delta_x = \imunit k$.
Using \eqref{eq:Qr_const} one finds
%
%\begin{subequations}
\begin{align}
\dJr &\equiv\ddiff \Jr \Al = \my \dHdAl -  \br{\rho\, \Ur^2}\m \label{eq:dJr}
%\dJr &\equiv\ddiff \Jr \Al = \my \dHdAl -  \frac{\rho\l \Ulr^2}{\Al} -  \frac{\rho\g \Ugr^2}{\Ag}\label{eq:dJr}
\intertext{(see \eqref{eq:asterisk_operator}) and}
%\dS &\equiv\ddiff \S \Al = \SAl-\SAg + \cnu \br {\SQl-\SQg}. \label{eq:dS}
\dS &\equiv\ddiff \S \Al = \SAl + \cnu \br {\SQl-\SQg}. \label{eq:dS}
\end{align}%
%\end{subequations}%
%
%A useful operator
%\[
%\psi\m = \frac{\psi\l}\Al+\frac{\psi\g}\Ag
%\]
%has here been introduced along with
%%
%Here, 
The source has here been parameterised as function of $\Al$ and $\Q\k$ with
\begin{align*}
%\S' &= \ddiff \S\Al, & 
\SAl = &\br{\pdiff \mcS\Al}_{\!\!\Ql,\Qg} \hspace{-.7cm},&
%\SAg = &\br{\pdiff \mcS\Ag}_{\!\!\Al,\Ql,\Qg}, \\
\SQl = &\br{\pdiff \mcS\Ql}_{\!\!\Al, \Qg} \hspace{-.7cm}, &
\SQg = &\br{\pdiff \mcS\Qg}_{\!\!\Al, \Ql} \hspace{-.7cm},  \hspace{.2cm}
\end{align*}
easily computed for any source $\mcS$ from discrete state differentials.

%\\
Some alternative forms of presenting $\dJr$ should also be pointed out, namely
\begin{equation}
\dJr = \frac{\symkappa^2-\br{\br{\rho\,\Ur}\m}^2}{\rho\m}
= -\rho\m \det\br{\Jacr}
= -\rho\m \lambda\_r^+\lambda\_r^-.
\label{eq:dJr_alternative}
\end{equation}

Extracting any particular growth rate or wave celerity from \eqref{eq:disp_eq} is perfectly straight forward and yields
\begin{equation}
\cnu = b_1 \pm \sqrt{b_1^2-b_0}
%{\cnu}_{,k,\p} = b_1 \pm_\p \sqrt{b_1^2-b_0}
\label{eq:disp_eq_cnu_extracted}
\end{equation}
with
\begin{align*}
b_1 &=\frac1{\rho\m}\br{ \br{\rho\, U}\m - \h\frac{\SQl-\SQg}{\delta_x} }, &
%b_1 &=\frac{ \br{\rho\, U}\m }{\rho\m}+ \h\frac{\SQg-\SQl}{\rho\m \ddisc_x},\\
%b_0 &= \br{\SAl-\J'}/\rho\m%\frac1{\rho\m}
%b_0 &= \frac{\SAl-\J'}{\rho\m}
%b_0 &= \frac1{\rho\m}\br{\frac{\SAl}{\delta_x}-\sqbrac{ \my \dHdAl -  \br{\rho\, U^2}\m}},
b_0 &= \frac1{\rho\m}\br{\frac{\SAl}{\delta_x}-\dJ},
\end{align*}
%
%and $c=\cnu-\imunit k\nu$.
and the definitions \eqref{eq:cnu}, \eqref{eq:delta} and \eqref{eq:bvt}.
The wave resulting from plus in \eqref{eq:disp_eq_cnu_extracted} will in the following be termed the `fast wave'. % and indexed $\p=1$. 
Conversely, the minus wave will be termed the `slow wave'. % and indexed $\p=2$.

In the case where $\nu = 0$, the marginal stability condition $c_k \in \mathbb R$, often called the VKH criterion,  has a particularly simple solution. Both $\dJr$ and $\dS$ are real in this case, so that \eqref{eq:disp_eq} boils down to 
\begin{align}
\dS = 0, 
\qquad
\dJr = 0,
\label{eq:VKH_crit}
\end{align}
with $\S = 0$ from the holdup equation \eqref{eq:holdup_eq}.
We may therefore regard the VKH criterion as the equilibrium state with respect to changes in phase fraction in the frame of a moving wave perturbation.
$\dS = 0$ then gives the critical wave celerity $c\_{crit}$ and wave growth will occur if $\dJr\of{c\_{crit}}<0$.

%To determine the actual wave growth, or the stability of a viscous system, equation~\ref{eq:disp_eq} must be solved for $\cnu = c+\imunit k\nu$, where $c$ is related to the disturbance evolution through \eqref{eq:bvt}.
%\\

Note that the rate of growth will in \eqref{eq:disp_eq} depend upon the wavenumber $k$ (present in $\delta_x$,) but that the condition for marginal stability, \eqref{eq:VKH_crit}, will not.

These results are identical to those provided in \egg, \cite{Barnea_VKH_and_IKH, Holmaas_roll_wave_model,Liao_von_Neumann}, though a different approach has been chosen which provides a physical interpretation.

\subsection*{IKH}
Let us conclude this section by remarking on some features of the so-called \textit{inviscid Kelvin-Helmholtz} stability criterion (IKH.) 
This is the stability of the two-fluid model \eqref{eq:base_model} without the source term; $\s \equiv 0$. 
\textcolor{\markercolor}{
From \eqref{eq:dJr_alternative} we do however see that celerity $c$ must turn complex if the eigenvalues  do.
The IKH criterion is thus really a test on hyperbolicity.
Inspecting the eigenvalues \eqref{eq:eigenvalues}, the `\textit{inviscid} Kelvin-Helmholtz' (IKH) criterion can thus be written 
\[\text{IKH neutral stability:}\qquad \symkappa = 0.\qquad\qquad\]
From \eqref{eq:dJr_alternative} we then find the `inviscid' critical celerity %$c\_{iv}$
\[
c\_{iv,crit} = \lambda^+\_{iv,crit} = \lambda^-\_{iv,crit} = \br{\rho u}\m/\rho\m.
\]
}
%Solving \eqref{eq:VKH_crit} we then find the `inviscid' celerity %$c\_{iv}$
%\[
%c\_{iv,crit} = \lambda^+\_{iv,crit} = \lambda^-\_{iv,crit} = \br{\rho u}\m/\rho\m.
%\]
%$\lambda^\pm$, from \eqref{eq:eigenvalues}, are the eigenvalues of $\pdiff \bf\bv$. 
%This relationship is also found by diagonalising \eqref{eq:base_model} directly.
%So, wave growth occurs in the frictionless model when $\lambda^\pm$ turns complex, \ie\ when $\symkappa^2<0$ and the system turns elliptic. 
%The IKH criterion is thus really a test on hyperbolicity. 

\begin{figure}[h!ptb]%
\includegraphics[width=\columnwidth]{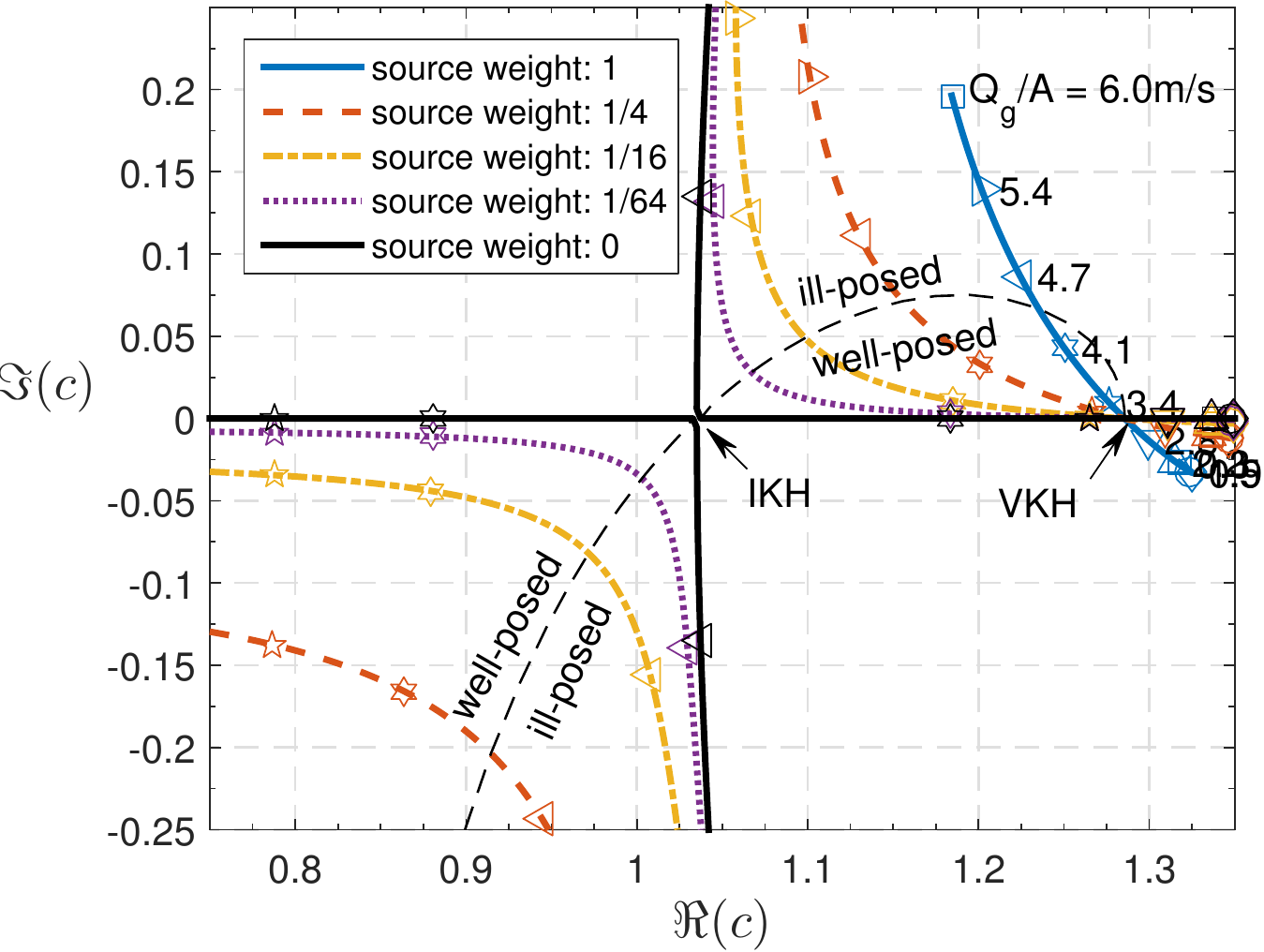}%
\caption{Complex celerity $c$ with altering superficial gas velocity $\USG$. $\Ql$, $\Al$ and the source differentials are kept constant about the critical VKH state. The source differentials are reduced in steps with a constant `source weight' between plots. Parameter values are presented in Section~\ref{sec:test_case} }%
\label{fig:IKH_vs_VKH}%
\end{figure}

Notice that the condition for IKH marginal stability, $\symkappa = 0$, does \textit{not} coincide with the VKH criterion \eqref{eq:VKH_crit} in the inviscid limit $\S\rightarrow0$.
This feature is illustrated in Figure~\ref{fig:IKH_vs_VKH}, showing $c$ in the complex plane with the parameters and closures described later in Section~\ref{sec:test_case}. For clarity, only the superficial gas velocity $\USG$ is altered and the source $\S$ is reduced sequentially towards zero by multiplying it by constant weights. 
This does not change the critical state, as long as $\S\not\equiv0$, but the rate of growth near the point of marginal VKH stability converges towards zero.
Figure~\ref{fig:IKH_vs_VKH} also shows the limit where the two-fluid model turns elliptical and ill-posed (assumed well-posed otherwise.)
There is a region of positive wave growth within which the viscous model remains hyperbolic. No such region exists in the inviscid model.

\section{Stability of Discrete Representations}
\label{sec:disc_stab}
We start the discrete analysis by examining the stability of representations of the two-equation model \eqref{eq:base_model}.
The remarks that then follow relates these results to representations of the four-equation model \eqref{eq:base_model_with_pressure}.
%Assume some general discrete differentiation operators represe
Let $(\delta \psi)_j\n $ symbolize the discrete differentiation operations used in the discretization to represent the partial differentials. 
%as $\pp \psi \of{x_j,t_n} \rightarrow (\delta \psi)_j\n$.
%$\partial_t \rightarrow \delta_t$, $\partial_x \rightarrow \delta_x$, $\partial_{xx} \rightarrow \delta_{xx}$.
System  \eqref{eq:base_model} may be written
\begin{equation}
%(\delta_t \bv)_j\n + (\delta_x \bf)_j\n  = \nu(\delta_{xx} \bv)_j\n + \bs_j\n.
(\delta_t \bv)_j\n + (\delta_x \bf)_j  = \nu(\delta_{xx} \bv)_j + \bs_j
\label{eq:discrete_base}
\end{equation}
after discretization.
%
%\begin{equation}
%\mathbb A_j\n \sqbrac{(\delta_t \bw)_j\n + \mathbb B_j^n(\delta_x \bm g)_j\n  - \nu(\delta_{xx} \bw)_j\n - \bm r_j\n} = 0.
%\label{eq:discrete_base_messy}
%\end{equation}
%
%A rather general discretization form is here show with the intention of demonstrating that the linear stability analysis on a non-staggered grid is insensitive to the choices of state variables and chain rule extensions.
%%A rather general discretization form is here shown with to allow for free choices when it comes to state variables and discretization.
%The intension of demonstrating that the linear stability analysis on a non-staggered grid is insensitive to these choices.
 $\nu$ is here whatever artificial numerical viscosity one chooses to impose on the discretization. For example, $\nu = \dx^2/2\dt$ and central differences for the spatial derivatives constitutes a Lax-Friedrich scheme. 
\\

\begin{table*}%[H]%[h!ptb]%
\centering
\begin{tabular}{|c|c||c|c|c|c|}
\hline
			&			&			&	$(\delta \psi)_j\n$			  & $\ddisc $  \\\hline
$\delta_t$	&$-\imunit kc$		&explicit	&  $\frac{1}{\dt}\br{  \psi_j\nn-\psi_j\n} + (\cdots)\n$ 	& $\frac1\dt\br{\eEuler^{-\imunit  \phit}-1}$ \\\hline
$\delta_t$ 	&$-\imunit kc$		&implicit	&  $\frac{1}{\dt}\br{  \psi_j\nn-\psi_j\n}+ (\cdots)\nn$ 	& $\frac 1\dt\br{1-\eEuler^{\imunit  \phit}}$ \\\hline
%$\delta_t$	&$-\imunit kc$		&Crank-Nicolson ($r=\tfrac12$)	& $\frac{1}{\dt}\br{  \psi_j\nn-\psi_j\n}+ r(\cdots)\nn+(1-r)(\cdots)\n$ 	& 
%$\frac 1\dt\frac{\eEuler^{-\imunit  \phit}-1}{r\br{\eEuler^{-\imunit  \phit}-1}+1}$ \\\hline
$\delta_t$	&$-\imunit kc$		&Crank-Nicolson & $\frac{1}{\dt}\br{  \psi_j\nn-\psi_j\n}+ \tfrac12(\cdots)\nn+\tfrac12(\cdots)\n$ 	& 
%$\frac 2\dt\frac{1-\eEuler^{\imunit  \phit}}{1+\eEuler^{\imunit  \phit}}$ \\\hline
$\frac {2i}\dt\tan\tfrac\phit2$ \\\hline
$\delta_x$	&$\imunit k$		&upwind & $\frac1\dx\br{ \psi_j-\psi\jm}$ 		& %$\frac1\dx\br{1-\eEuler^{-\imunit \phix}}$ \\\hline 
 											$ \frac{2i}{\dx} \sin{\tfrac\phix2} \, \eEuler^{-\imunit \frac\phix2} $\\\hline 
$\delta_x$	&$\imunit k$		&central difference& $\frac1{2\dx}\br{ \psi\jp-\psi\jm}$ 	& $\frac \imunit \dx \sin\phix$\\\hline
$\delta_x$	&$\imunit k$		&staggered& $\frac1{\dx}\br{ \psi\jph-\psi\jmh}$ 	& $\frac {2i}\dx \sin{\tfrac\phix2}$\\\hline
$\delta_{xx}$ &$-k^2$	&central difference& $\frac1{\dx^2}\br{ \psi\jp-2\psi_j+\psi\jm}$ 	& $\frac 2{\dx^2}\br{\cos\phix-1}$\\\hline
$\ol {\psi_j}$ &$\Psi$	&mean	& $\frac12\br{\psi\jph+\psi\jmh}$ 			& $ \cos\frac\phix2 $ \\\hline
\end{tabular}
\caption{Some discrete differentiation operators and their corresponding wave operator. $\phix = k \dx$ and $\phit = k c \dt$.}
\label{tab:delta}
\end{table*}

First, let us regard the stability of \eqref{eq:discrete_base} in the context of a common von Neumann analysis \cite{von_Neumann_original}.  
Introduce $\bvt_j\n = \bv_j\n - \bV$
and express $\{\bvt_j\n\}$ and $\{\bvt_j\nn\}$ with spatial Fourier transforms
\begin{equation}
%\bv_j^m = \sum_k \bvh_k^m \eEuler^{\imunit kx_j}, \qquad m = n, n+1.
\bvt_j\n = \sum_k \bvh_k\n \eEuler^{\imunit kx_j}, \qquad \bvt_j\nn = \sum_k \bvh_k\nn \eEuler^{\imunit kx_j}.
\label{eq:bvj_Ftransform}
\end{equation}
Insertion into \eqref{eq:discrete_base} and dropping higher order terms yields a system on the form 
\[
\sum_k \br{ \bvh_k\nn - \G_k \bvh_k\n }\eEuler^{\imunit kx_j} = \bm 0,
\]
$\G$ being the amplification matrix. We have
\begin{equation}
\bvh_k\nn = \G_k\bvh_k\n = \G_k\G_k\bvh_k\nm = (\G_k)\nn \bvh_k^0, 
\label{eq:Gnn}
\end{equation}
which provides the result of a common von Neumann analysis, namely that the spectral radius of $\G_k$ must be less or equal to one as a necessary condition for system stability.
\\

Rather than examining $\G_k$, we will use the continuous stability analysis from the previous section for which the solution is already calculated.
Notice that the equation~\eqref{eq:bvj_Ftransform}, once \eqref{eq:Gnn} is inserted, can be written on exactly the form \eqref{eq:general_sol_lin_syst}, provided $\G_k$ is diagonalisable. 
In this case,
$\PP$ contains the eigenvectors of $\G_k$ and $\cdisc_{k,\p} = \frac{\imunit }{k}\frac{\ln \lambda_{\mr G_k,\p}}{\dt}$, $\lambda_{\mr G_k,\p}$ being the $\p$-th eigenvalue of $\G_k$.
In fact, this is a solution at any point in our discrete system for as long as non-linear effects remain negligible.
Again, these are just linear combinations of modes; we may express the  discrete point solution analogous to \eqref{eq:bvt_disc} by
\begin{equation}
%\bv_j\n = \sum_\p\bvh_{\p} \eEuler^{ik_\p(x_j-\cdisc_{\p}t_n)}. 
\bvt_j\n = \sum_k \sum_{\p=1,2} \bvh_{k,\p} \eEuler^{\imunit k(x_j-\cdisc_{k,\p}t_n)}.
\label{eq:bvt_disc}
\end{equation}
Inserting \eqref{eq:bvt_disc} into the discrete model \eqref{eq:discrete_base} and linearizing yields a system analogous to \eqref{eq:base_model_linearized_dEdV},
\begin{equation}
%\pdiff\bE \bV \bvt = 0
%\sum_\p\pdiff{\bEdisc_\p} \bV \bvh_\p  \eEuler^{ik_\p(x_j-\cdisc_\p t_n)} = 0,
\sum_k \sum_{p=1,2}\pdiff{\bEdisc_{k,\p}} \bV \bvh_{k,\p}  \eEuler^{\imunit k(x_j-\cdisc_{k,\p} t_n)} = \bm 0,
\label{eq:base_model_linearized_dEdV_disc}
\end{equation}
$\bEdisc$ being the discrete equivalent of $\bE$ from \eqref{eq:E}, differing only in that discrete differential operators 
\begin{equation}
\ddisc \equiv \frac{\br{\delta\exp \imunit k (x-\cdisc\,t)}_j\n}{\exp \imunit k (x_j-\cdisc\,t_n)}
%\ddisc \equiv \frac{\br{\delta \,\eEuler^{ \imunit k (x-\cdisc\,t)}}_j\n}{\eEuler^{ \imunit k (x_j-\cdisc\,t_n)}}
\label{eq:ddisc}
\end{equation}
replace $\delta$. 
These $\ddisc$ terms, approximating $\delta$, hold all numerical error in its entirety and are simple algebraic expressions. 
Again, $\bEdisc$ is independent of $n$ and $j$ so that each $k,\p$-term must equal zero individually. 
The problem is now equivalent to \eqref{eq:base_model_dEdV_nosum} and its solution is obtained directly from \eqref{eq:disp_eq}-\eqref{eq:dS}, with two celerities for each wavenumber $k$. 
Also the discrete viscous celerity 
%\[
%\cnudisc \equiv - \frac{\ddisc_t}{\ddisc_x} + \nu \frac{\ddisc_{xx}}{\ddisc_x}
%\]
$\cnudisc $
follow the definition \eqref{eq:cnu} using the \textit{discrete} $\ddisc$ operators.
$\ddisc$  gives the value of $\ddisc_t$, which in turn gives the wave growth and dispersion from the chosen time discretization method.
For example, the explicit or \textit{forward Euler} time integration has the operator $\ddisc_t=\frac 1\dt\br{\eEuler^{-\imunit  k \cdisc}-1}$.
Solving for $\cdisc$ yields $\cdisc = \tfrac1{k\dt} \ln\br{1+\ddisc_t\dt}$
with $\ddisc_t = \nu\ddisc_{xx}-\cnudisc\ddisc_x$ from \eqref{eq:cnu}.%
\footnote{
We may generalize the discrete time differential operators listed in Table~\ref{tab:delta} by introducing
 the `degree of implicitness' $r$ as a linear combination of the forwards and backwards Euler integrations, 
$\tfrac1\dt\br{\psi\nn-\psi\n}+r(\cdots)\nn+(1-r)(\cdots)\n$.
%More generally, if we introduce a weight $r$ representing the `degree of implicitness,' time discretization $\tfrac1\dt\br{\psi\nn-\psi\n}+r(\cdots)\nn+(1-r)(\cdots)\n$ has 
The discrete differential operator for this integration is $\ddisc_t=\frac 1\dt\frac{\exp\br{-\imunit  k \cdisc}-1}{r\br{\exp\br{-\imunit  k \cdisc}-1}+1}$,
giving $c = \tfrac1{k\dt} \ln\br{\frac{1+(1-r)\ddisc_t\dt}{1-r\ddisc_t\dt}}$.
}

%For example, the more general time discretization $\tfrac1\dt\br{\psi\nn-\psi\n}+r(\cdots)\nn+(1-r)(\cdots)\n$ has the discrete differential operator $\ddisc_t=\frac 1\dt\frac{\eEuler^{-\imunit  k c}-1}{r\br{\eEuler^{-\imunit  k c}-1}+1}$. Solving for $c$ yields 
%$c = \tfrac1{k\dt} \ln\br{\frac{1+(1-r)\ddisc_t\dt}{1-r\ddisc_t\dt}}$
%with $\ddisc_t = \nu\ddisc_{xx}-\cnudisc\ddisc_x$ from \eqref{eq:cnu}.
%%These give in turn the wave growth and dispersion through the time discretization $\ddisc_t$.
$\ddisc$-functions for some of the most common discrete differentiations are presented in Table~\ref{tab:delta}. 
Tabulated operators are presented with phase angles $\phix = k \dx$ and $\phit = k \cdisc \dt$, which represent the phase rotation within a grid cell length or time step, respectively.
Constructing similar operators for more complicated interpolations is usually straight forward. For instance, $\ddisc_x$ in the QUICK scheme is $\frac{1}{8\dx}\br{3e^{\imunit \phix}+3-7e^{-\imunit \phix}+\eEuler^{-i2\phix}}$.
Notice that all operators listed in Table~\ref{tab:delta} are consistent, \ie, they satisfy $\ddisc \rightarrow \delta$ as $\dx$ and $\dt$ approach zero.

\begin{remark}
\label{rem:convergence}
The stability behaviour of a discrete representation will converge towards that of the continuous model if 
\[
\cnudisc  \equiv - \frac{\ddisc_t-\nu\ddisc_{xx} }{\ddisc_x}  \rightarrow c + \imunit k\nu.
\] 
For this to happen, all $\ddisc$ operators must be consistent ($\ddisc\rightarrow \delta$) and 
$\dx$ and $\dt$ must approach zero together, smoothly.
\end{remark}

%\begin{remark}
%\label{rem:variable_extension_independence}
%The predicted linear stability is independent of the choice of discrete variables in \eqref{eq:bvj_Ftransform}.
%\end{remark}

\begin{remark}
\label{rem:variable_independence}
The predicted linear stability is independent of which variable the discrete system is solved for, provided the discrete differentiations $\ddisc$ are independent of this choice.
\end{remark}

\begin{remark}
\label{rem:system_form}
The predicted linear stability is independent of the form of the discrete system, be it conservative, primitive, four-equation, two-equation, etc., provided the discrete differentiations $\ddisc$ are independent of these choices.
% and compressibility effects are negligible.
\end{remark}

%\begin{remark}
%Incompressible fluids are assumed in the stability analysis, even though we are representations of the compressible model. 
%This is common practise, but denies us information about the sonic stability.
%\end{remark}

Remark~\ref{rem:variable_independence} and \ref{rem:system_form} are results of the linearisation. 
This is briefly illustrated by Taylor expanding any chosen variable $\bv\of\bw$ about $\bW$ which yields
%\begin{proof}
%Taylor expanding any chosen variable $\bv\of\bw$ about $\bW$ yields
\begin{equation*}
\bvt = \bV + \pdiff{\bV}{\bW}\bwt + \mc O\of{|\bwt|^2},
%\label{eq:}
\end{equation*}
where $\bV = \bv \of \bW$.
% and $\pdiff{\bV}{\bW}$ is the Jacobian of $\bV$ evaluated at $\bW$.
%Provided an invertible Jacobian exists 
We may thus define 
\begin{equation*}
%\bvt \equiv \pdiff{\bV}{\bW}\bwt.
\bwt \equiv \pdiff{\bW}{\bV}\bvt
%\label{eq:}
\end{equation*}
and freely impose variable transformations which does not affect the \textit{linear} stability of the system, provided those transformation matrices exist.
%This proves Remark~\ref{rem:variable_extension_independence}. 

Further, in linearizing an arbitrary discretized variable $\bw$, and remembering that $\br{\delta\bW}_j\n=0$, one finds
\[
(\delta \bw)_j\n \rightarrow \ddisc \bwt = \pdiff{\bW}{\bV} \ddisc \bvt.
\]
The linear system of one variable will thus only be a factorization of the same system in another variable,
as alluded to in Remark~\ref{rem:variable_independence}. 
Also noted in the remark, $\ddisc$ must be unchanged in the variable transformation of the last equality.

Finally, any type term representation based on the chain rule will, after the linearization, be equivalent. 
For example
\[
\brt{ \pdiff f w }_j\n\br{\delta \bw}_j\n \rightarrow \br{\pdiff \bF\bW + \mc O\of{\bwt}} \ddisc \bwt = \ddisc \bft.
\]
Any discrete representation of system~\ref{eq:base_model_with_pressure} will thus be equivalent, provided the discrete differential operators $\ddisc_t$, $\ddisc_x$ and $\ddisc_{xx}$ are respectively the same. 
%\end{proof}
\\

A consequence of Remark~\ref{rem:system_form} is that four equation formulations of the compressible system~\eqref{eq:base_model_with_pressure}
is equivalent to \eqref{eq:discrete_base} provided we do not mix different discrete differentiations. 
Examples of commonly used mixed discrete differentiations are convection flux terms of the form 
\begin{equation*}
\br{\psi u}\jph =  
\begin{cases}
\psi_j\ol{ u_{j+\frac12}} & \text{if}~~ \ol{ u_{j+\frac12} }\geq 0,\\
\psi\jp\ol{ u_{j+\frac12}} & \text{otherwise},
\end{cases}
\end{equation*} 
(which linearizes to a central difference in $\wt u$ and an upwind difference in $\wt \psi$.) Staggered grid formulations, in which the differentiation depends on the proximity of the data points, is another example.

%However, 
Incompressible fluids are assumed in the stability analysis itself in the present work, even though the representation is of the compressible model. 
This is common practise and provides us with simple, explicit stability expressions, though denies us information about the sonic stability. 
Including density variations is straightforward, but necessitates solving a higher order dispersion equation numerically.

\begin{example}[The stability of a Lax-Friedrich and a local Lax-Friedrich scheme]
\label{ex:LLF}
%Consider a Roe scheme as presented in \ref{sec:Roe_scheme}.
The Lax-Friedrich scheme is commonly written
\[
\frac{\bv\nn-\bv\n}{\dt} + \frac{\bf\jph-\bf\jmh}{\dx} = \bs_j
\]
with
\[
%	\bf\jph = \tfrac12\br{\bf\jp + \bf_j} - \tfrac12\max |\lambda^\pm|\br{\bv\jp-\bv_j},
	\bf\jph = \tfrac12\br{\bf\jp + \bf_j} - \frac \nu\dx\,\br{\bv\jp-\bv_j}.
\]
Numerical viscosities in the Lax-Friedrich scheme and the local Lax-Friedrich scheme are then
\begin{align*}
\nu\^{LF} &= \frac{\dx^2}{2\dt}
& &\text{and} &
\nu\^{LLF} &= \frac{\dx}{2}\max |\lambda^\pm|,
\end{align*}
respectively. 
%
%In the case of a common Lax-Friedrich scheme 
%\[
%\nu^{LF} = \frac{\dx^2}{2\dt}
%\]
%and in case of a local Lax-Friedrich scheme
%\[
%\nu^{LLF} = \frac{\dx}{2}\max |\lambda^\pm|.
%\]
Thus, the local Lax-Friedrich scheme determines the artificial numerical diffusion according to spectral radius of $\jac$.
%\ie, the local spectral radius of $\jac$ determines the artificial numerical diffusion. 
Eigenvalues, presented in \eqref{eq:eigenvalues}, are % and we have that $\max |\lambda^\pm| = (|(\rho u)\m| + \symkappa)/\rho\m$,
%$\max |\lambda^\pm|$ is here computed form the steady state $\bV$.
computed at the steady state $\bv = \bV$.

These schemes use simple central differences; from Table~\ref{tab:delta} we find $\ddisc_x = \frac \imunit \dx \sin\phix$ and $ \ddisc_{xx} = \frac2{\dx^2}\br{\cos\phix-1}$.
Extracting the stability equations of either scheme is strikingly easy;
equation \eqref{eq:disp_eq_cnu_extracted} gives $\cnudisc$, $\ddisc_t$ is obtained from the definition \eqref{eq:cnu} and the complex wave celerity from the expression of the chosen time discretization.

%Extracting the stability equations of either scheme is strikingly easy. 
%After linearization, it takes the form of 
%\eqref{eq:base_model_linearized} with 
%\begin{align*}
%%\ddisc_t &= \frac{1-\eEuler^{-\imunit  \phit}}\dt
%%&
%\ddisc_x &= \imunit  \frac {\sin\phix}\dx,
%&
%\ddisc_{xx} &= 2\frac {\cos\phix-1}{\dx^2}
%%&
%% \nu &= \frac{\dx}{2}\max |\lambda^\pm|, 
%\end{align*}
%%
%and whatever time discretization is used.
%Equation \eqref{eq:disp_eq_cnu_extracted} 
%once more give the wave speed and wave growth. 
\end{example}

%\begin{example}[The stability of a Roe scheme]
%\label{ex:Roe}
%Consider a Roe scheme as presented in \ref{sec:Roe_scheme}.
%Extracting the stability equations is strikingly easy. 
%After inserting \eqref{eq:Roe_sol_f} for $\bf\jph$ and $\bf\jmh$ and linearizing, it takes the form of 
%\eqref{eq:base_model_linearized} with 
%\begin{align*}
%%\ddisc_t &= \frac{1-\eEuler^{-\imunit  \phit}}\dt
%%&
%\ddisc_x &= \imunit  \frac {\sin\phix}\dx 
%&
%\ddisc_{xx} &= 2\frac {\cos\phix-1}{\dx^2},
%&
% \nu &= \frac{\dx}{2} \LL^{\text{-} 1}|\Lamb| \LL, 
%\end{align*}
%%
%and whatever time discretization is used.
%The diagonalisation matrices in $\nu$ are computed form the steady state $\bV$. % with $\frac{h\jp-h_j}{\a_{\ell,j+1}-\a_{\ell,j}}\rightarrow \dHdAl$.
%%Equations \eqref{eq:disp_eq}-\eqref{eq:dS} 
%Equation \eqref{eq:disp_eq_cnu_extracted} 
%once more give the wave speed and wave growth. 
%\end{example}

\begin{example}[The stability of a scheme for the compressible model on a staggered grid]
\label{ex:compressible}

Say we wish to simulate the compressible model \eqref{eq:base_model_with_pressure} with a staggered grid discretizations.
%Compressible schemes may be based on the model \eqref{eq:base_model_with_pressure}.
Staggered grids are quite common with this model as the staggered grid offers a tight stencil for most of the data points and denies so-called \textit{checkerboard} solutions of the pressure field (see \egg\ \cite{Ferziger_book_basic_CFD}.)

%Staggered grids are well-known in computational fluid dynamics (see \egg\ \cite{Ferziger_book_basic_CFD}.)
Upper-case indices $J$ are used for the mass control volume centre points, and lower-case indices $j$ for the momentum ones.
%The staggered grid is constructed with the mass control volumes shifted spatially half a cell length in front of the momentum control volumes, \ie, $x_J = x_j+\h \dx$, as illustrated in Figure~\ref{fig:staggered_grid}. 
The staggered grid is constructed with the momentum control volumes shifted spatially half a cell length behind the mass control volumes, \ie, $x_j = x_J-\h \dx$, as illustrated in Figure~\ref{fig:staggered_grid}. 

\begin{figure}[h!ptb]%
\centering
\includegraphics[width=.8\columnwidth]{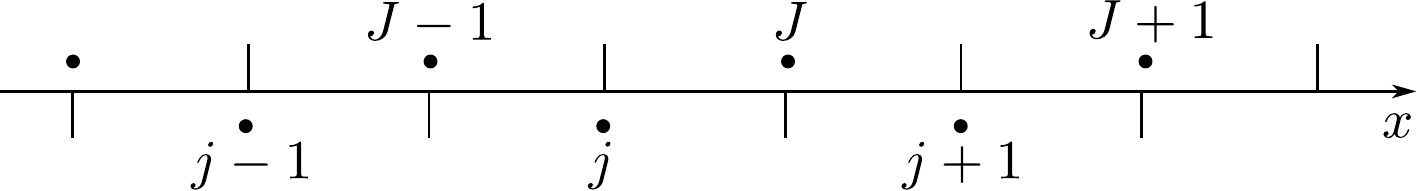}%
\caption{The staggered grid; cells and indices.}%
\label{fig:staggered_grid}%
\end{figure}

As opposed to on a co-located grid (non-staggered,) the orientation of the data points is likely to affect the way in which individual terms are discretized, and therefore also the stability behaviour. 
A particular choice has therefore been made for the variables of this example, namely physically conserved variables of specific mass $\mm\k = \rho\k a\k$ and specific momentum  $\ii\k = \rho\k \q\k$.
A sensible discretization can be written
\begin{align*}
&(\delta_t \mm\k)_J\n + (\delta_x \ii\k)_J\n = \nu (\delta_{xx} \mm\k)_J\n 
\\
&
\begin{aligned}[t]
(\delta_t \ii\k)_j\n +(\delta_x ( \ii u)\k)_j\n  % \brt{\delta_x  \frac{\imunit \k^2}{m\k}}_j\n % 
+ \ol{\a\kj} (\delta_x p)_j\n  + g_y \ol{\mm\kj}  (\delta_x h)_j\n&   
\\=   \nu (\delta_{xx} \ii\k)_j\n +s\kj,&
\end{aligned}
\\
&\a_{\ell,J}+ \a_{\mr g,J} = \area, \quad \a\kJ = \frac{\mm\kJ}{ \rho\kJ},
\end{align*} 
with
\begin{align*}
p_{J} &= \mc P\of{\mm_{\ell,J},\mm_{\mr g,J}}, \\
h_J &= \mc H\of{\a_{\ell,J}},\\
s\kj &= \mc S\k\of{\ol{\a\kj},\frac{\ii\lj\n}{\ol{\rho\lj}},\frac{\ii\gj\n}{\ol{\rho\gj}}},
\end{align*} 
and arithmetic averaging
\[
\ol{w_j} = \tfrac12\br{w_J + w_{J-1}}
\]
being used in between data points.
%Which discretization best suits the the momentum convection term $(\delta_x ( \ii u)\k)_j\n $ is up for discussion and is for now kept purely symbolic. %generic.
The momentum convection term $(\delta_x ( \ii u)\k)_j\n $ is for now kept purely symbolic. %generic.
Compressibility effects are dismissed in the linear analysis and densities again made constant, as in all VKH analyses.
%The discrete pressure differential $(\delta_x p)_j\n$ can be eliminated between the momentum equations by subtracting the gas equation from the liquid equation after dividing the respective equations by $\ol{\a\kj}$, 
The discrete pressure differential $(\delta_x p)_j\n$ can again be eliminated between the momentum equations after dividing eace by $\ol{\a\kj}$. %,
%similar to the derivation of the incompressible model \eqref{eq:base_model}.
Thus, a Fourier solution on the form \eqref{eq:bvt_disc} once more yields a linearized system similar to \eqref{eq:base_model_linearized_dEdV_disc},
but with the variables
$
%\bwt = \br{\tilde m\l,\tilde m\g,\tilde \ii\l, \tilde \ii\g}^T
\bwt = \br{\tilde a\l,\tilde a\g,\tilde q\l, \tilde q\g}^T
$.
The system reads
\begin{equation*}
\pdiff {\bE\^d} \bW \bwh=
\begin{pmatrix}
-\cnudisc\,\wh a\l + \delta_x \wh q\l 
\\
-\cnudisc\,\wh a\g + \delta_x \wh q\g 
\\
\sqbrac{
\begin{gathered}%[c]
\diffk{\frac{\rho}{\A}\br{-\cnudisc\,\wh\q +\widehat{\ddisc_x \q u}}} + \my \dHdAl\ddisc_x \wh\a\l 
\\ - \wh\a\l \cos\! \tfrac\phix2\, \SAl  - \wh\q\l\SQl  - \wh\q\g\SQg %
\end{gathered}}
\\
\wh\a\l+\wh \a\g
\end{pmatrix}
\end{equation*} 
%
%\begin{equation*}
%\pdiff {\bE\^d} \bW \bwh=
%\begin{pmatrix}
%-\cnudisc\,\wh a\l + \delta_x \wh q\l 
%\\
%-\cnudisc\,\wh a\g + \delta_x \wh q\g 
%\\
%\sqbrac{
%\begin{gathered}%[c]
%\frac{\rho\l}{\A\l}\brt{-\cnudisc\,\wh\q +\widehat{\ddisc_x u\q}}\l-\frac{\rho\g}{\A\g}\brt{-\cnudisc\,\wh\q +\widehat{\ddisc_x u\q}}\g 
%\\ + \my \dHdAl\ddisc_x \wh\a\l  - \wh\a\l \cos\! \tfrac\phix2\, \SAl  - \wh\q\l\SQl  - \wh\q\g\SQg %
%\end{gathered}}
%\\
%\wh\a\l+\wh \a\g
%\end{pmatrix}.
%\end{equation*} 
%
with $\diffk\cdot=\br\cdot\l-\br\cdot\g$.
The tight stencil reduces the error in the $\ddisc_x$ operator to
$
\ddisc_x = \tfrac{2i}\dx \sin{\!\tfrac \phix2}
$
for those data points located favourably on the staggered grid.
%with the discrete viscous celerity $\cnudisc$ defined as in \eqref{eq:cnu} from this and the other discrete time and Laplacian operators.
In this case only the momentum convection term requires an alternative form of spacial differencing.
We split it
%The momentum convection term is now split 
into $\wh\a\k$ and $\wh \q\k$ components and define $\ddisc_{\a,x}$ and $\ddisc_{\q,x}$ by
\[
\widehat{\ddisc_x (\q u)}\k =  2 U\k \ddisc_{\q,x} \wh q\k  - U\k^2 \ddisc_{\a,x} \wh \a\k
\]
so that the term become analogues to the previous examples. 
%Notice that the form of the mass equations is identical to the previous systems, suggesting that we should again define a discrete viscous celerity
%\[
%\cnudisc = - \frac{\ddisc_t-\nu\ddisc_{xx}}{\ddisc_x}.
%\]

The simplest approach is 
to take the determinant of $\pdiff {\bE_\p\^d} \bW$, requiring it to be zero for every wavenumber. 
%The resulting dispersion equation is again quite similar to the continuous one, namely
%%
%\begin{equation*}
%\dJrd\, \ddisc_x  - \dSd = 0
%%\label{eq:disp_eq}
%\end{equation*}
%
A discrete version of \eqref{eq:disp_eq} then emerges
with
\begin{align*}
\dJrd &= \my \dHdAl - \brt{\rho\br{ U^2\tfrac{\ddisc_{\a,x}}{\ddisc_x} - 2 U\cnudisc \tfrac{\ddisc_{\q,x}}{\ddisc_x} + {\cnudisc}^2   }}\m,
\\ %\intertext{and}
\dSd &  = \cos\!\tfrac \phix2 \,\SAl + \cnudisc \br {\SQl-\SQg}.
\end{align*}%
$\dJrd$ is of course identical in form to \eqref{eq:dJr} if $\ddisc_{\a,x}=\ddisc_{\q,x}=\ddisc_{x}$, and $\dSd$ to \eqref{eq:dS} if not for the arithmetic average, as pointed out in Remark~\ref{rem:system_form}.

An upwind-type interpolation may be chosen for the momentum convection term to complete the example.
If we choose, say,
\[
(\delta_x ( \ii u)\k)_j\n = \tfrac 1\dx\br{ \mm\kJ \ol{u\kj}^2-\mm\kJm \ol{u\kjm}^2},
\]
with $\ol{u\kj} = \ii\kj/\ol{\a\kj}$,
we get
\begin{align*}
\ddisc_{\a,x} &= \tfrac{2i}{\dx}\br{\eEuler^{-\imunit \frac\phix2}\sin\phix-\sin\tfrac\phix2}, %\frac{2 \imunit }{\dx} \sin\frac\phi2,
\\
\ddisc_{\q,x} &= \tfrac{2i}{\dx} \eEuler^{-\imunit \frac\phix2}\sin\tfrac\phix2.
%\ddisc_{\q,x} &= \frac{1}{\dx} \br{1-\eEuler^{-\imunit \phi}}.
\end{align*}
This is a rather diffusive choice, made to stabilize the scheme as the staggered conservative variable formulation turns out to provide very little diffusion otherwise. Solving for other variables, such as phase fractions and velocities, is also quite common and would entail other differentiations and result in a different dispersion equation. 
One may for example look to the example in \cite{Liao_von_Neumann}, although this example appear to neglect that some of the  information in the momentum equation is dislocated.
\end{example}

\begin{example}[The Stability of a Roe Scheme]
\label{ex:Roe}
The Roe scheme is presented in \ref{sec:Roe_scheme}.
A viscous matrix, the Roe matrix, is used in this scheme in place of a scalar viscosity. Let's split it
into a diagonal part and an off-diagonal part:
\[
\tfrac\dx2\, \LL\inv |\Lamb| \LL = \nu \I + \nu\_T
\begin{pmatrix}
	0& 1/\symkappa\\ \symkappa & 0
\end{pmatrix}
\]
where
\begin{align*}
\nu &= \tfrac\dx2  \tfrac12\br{|\lambda^+|+|\lambda^-|}, 
&
\nu\_T &= \tfrac\dx2 \tfrac12\br{|\lambda^+|-|\lambda^-|}  
\end{align*}
Following the previous procedure, we regain the form \eqref{eq:disp_eq}, but with 
\begin{align*}
\dJrd &= \frac{\br{\gamma\symkappa}^2 - \br{\br{\rho\,\Ur}\m}^2 }{\rho\m} 
=  \my\dHdAl-\br{\rho\,\Ur^2}\m \!- \frac{1-\gamma^2}{\rho\m}\symkappa^2,\\
\dSd &= \gamma \SAl + \cdisc_{\nu,\nu\_T}\br{\SQl-\SQg},
\end{align*}
where
\begin{align*}
\gamma &= 1- \nu\_T  \frac{\rho\m}{\symkappa}\frac{\ddisc_{xx}}{\ddisc_x} 
%\intertext{and}
&
&\text{and}
&
\cdisc_{\nu,\nu\_T} &= \cnudisc -  \nu\_T \frac{ \br{\rho\,U}\m}{\symkappa}\frac{\ddisc_{xx}}{\ddisc_x}.
\end{align*}
Central difference operators $\ddisc_x = \tfrac \imunit \dx \sin\phi_x$, $\ddisc_{xx} = \tfrac 2{\dx^2}(\cos\phix-1)$ are again implied.
The viscous terms will disappear form $\cdisc_{\nu,\nu\_T}$ if both characteristics are of the same sign;
the predicted wave celerity will be very precise as the flow turns unstable.
\\

In fact, because  $\LL\inv |\Lamb| \LL$ equals the Roe matrix $\Big(\Jac\Big)\^{Roe}$ from \eqref{eq:Roe_problem} if both eigenvalues are positive, and $-\Big(\Jac\Big)\^{Roe}$ if both eigenvalues are negative, and because the Roe matrix is designed to obey
\[
\Big(\Jac\Big)\^{Roe}\jph(\bv\jp-\bv_j) = \bf\jp-\bf_j,
\]
the stability of the Roe scheme \eqref{eq:Roe_sol_f} is identical to that of the simple upwind scheme
\begin{equation}
\bf\jph = \begin{cases}
\bf_j, & \lambda^+> 0,\,\lambda^- > 0\\
\bf\jp, & \lambda^+< 0,\,\lambda^- < 0
\end{cases}
\label{eq:Roe_reduces_to}
\end{equation}
in the case of supercritical flow.
Then, the growth and celerity equations once more reduce to \eqref{eq:disp_eq_cnu_extracted} with the upwind differentiation $\ddisc_x = \frac1\dx\br{1-\eEuler^{-\imunit \phix}} = \frac{2i}{\dx} \sin{\tfrac\phix2} \, \eEuler^{-\imunit \frac\phix2}$ and no net artificial  viscosity; $\nu=0$.

\textcolor{\markercolor}{
Equation \eqref{eq:dJr_alternative} reveals that the VKH criterion \eqref{eq:VKH_crit}, where wave growth is at an equilibrium with $c = c\_{crit}\in \mathbb R$, 
coincides with hydraulically critical flow relative to the perturbation wave. 
That is, one of the relative eigenvalues $\lambda\_r^\pm = \lambda^\pm-c$ equals zero at neutral stability.
This means that the Roe scheme is equivalent to \eqref{eq:Roe_reduces_to} whenever the VKH wave growth is positive.
}

%At the VKH criterion \eqref{eq:VKH_crit}, where wave growth is at an equilibrium with $c = c\_{crit}\in \mathbb R$, 
%\eqref{eq:dJr_alternative} reveals that 
%eigenvalues are found to be
%$\lambda\_{crit}^+ = 2\frac{(\rho u)\m}{\rho\m}-c\_{crit} > \lambda\_{crit}^- = c\_{crit} >0$.
%This means that the Roe scheme is equivalent to \eqref{eq:Roe_reduces_to} whenever the VKH wave growth is positive.

\end{example}

\section{Numerical Tests and Results}
\label{sec:Num_test}
Predictions from a number of schemes will here be presented, namely the explicit and implicit variants of the Lax-Friedrich scheme (Example~\ref{ex:LLF}), abbreviated LF, the staggered upwind scheme solved for conservative variables (Example~\ref{ex:compressible}), abbreviated UWS, and the Roe scheme (Example~\ref{ex:Roe}.)
The aim of these comparisons is not to establish a favourite amongst the chosen representations, but to demonstrate how the linear theory provides a powerful simulation support tool. Indeed, multiple considerations are important when choosing a scheme. Many choices can be made both stable and accurate if the simulation parameters are collected with the aid of the hitherto presented linear theory.

\subsection{Initial Conditions} % for Single-Wave Simulations }
Disregarding compressibility, the flow development that springs out from the initial conditions will generally consist of two waves per wavenumber, as given in the solution \eqref{eq:bvt} or \eqref{eq:bvt_disc}. 
%This is because the system rank is 2 and so the two components of $\bvh_k^0$ offers two degrees of freedom; the solution 
These solutions show that $\bvh_k^0 = \sum_\p\bvh_{k,\p}$, each $\bvh_{k,\p}$ superimposing one of the two $c_{k,\p}$ waves. 
In order for a simulation to provide only a single wave $c_{k_1,\p_1}$ the initial conditions must be 
$\bvh_{k_1}^0 = \bvh_{k_1,\p_1}$, $\bvh_{k\neq k_1}^0=0$, where  $\bvh_{k_1,\p_1}$ satisfies \eqref{eq:base_model_dEdV_nosum}.
This was implicitly carried through in the VKH derivation of Section~\ref{sec:VKH} by the transformation $\pdiff \bV\Al \hat\a\l = \bvh$, which revealed \eqref{eq:Qr_const}.
%We further found \eqref{eq:Qr_const} which gives 
\eqref{eq:Qr_const} implies the transformation $\pdiff \bV\Al = \br{1,-\br{\rho\,\Ur}\m}^T$. Thus,
after choosing a volumetric disturbance $\hat\a_{\ell,k_1}$, a pure $c_{k_1,\p_1}$ disturbance wave is obtained by choosing 
\begin{equation}
\hat v_{2,k_1}^0 = \diffk{\rho \hat u}=-\br{\rho\br{U-c\^d_{\nu,k_1,\p_1}}}\m \, \hat\a_{\ell,k_1}^0.
\label{eq:num_exp:init_cond}
\end{equation}
Corresponding primitive variables $\hat u\k$ are found from the transformation matrix \eqref{eq:transformation_matrices:dwdv}.

\begin{figure}[h!ptb]%
\includegraphics[width=\columnwidth]{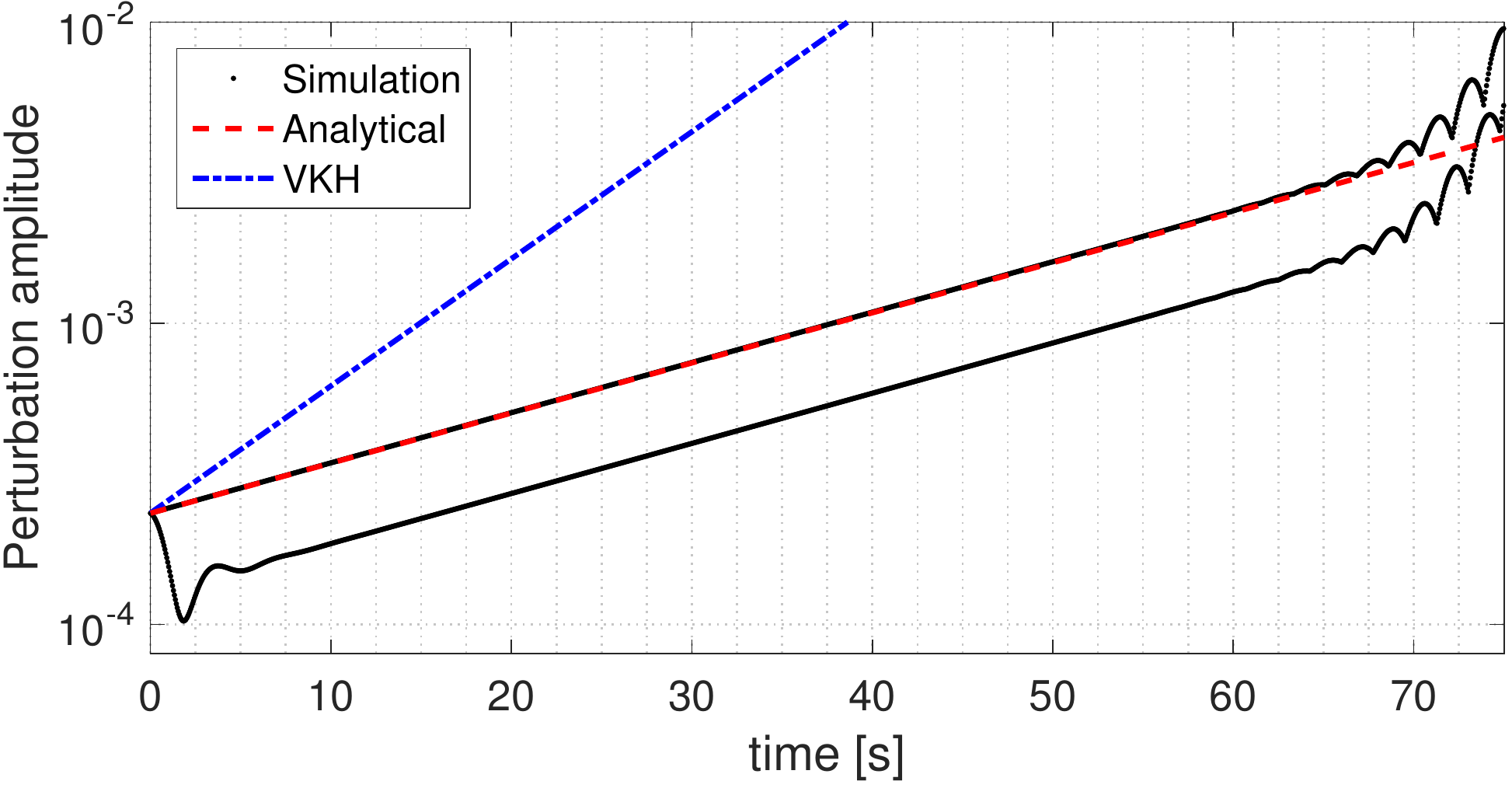}%
\caption{
\textcolor{\markercolor}{Perturbation wave amplitude $\max_j \{\tilde\a\lj\n\}/\area$ vs.\ time.}\\
Comparing single wave simulation with initial conditions $\bvh_{k_1}^0 = (\hat\a_{\ell,k_1}^0,0)^T$ (plot with initial disturbance) to simulation with initial conditions from \eqref{eq:num_exp:init_cond} (plot without initial disturbance); explicit Lax-Friedrich simulation.}%
\label{fig:num_exp:init_cond}%
\end{figure}

Figure~\ref{fig:num_exp:init_cond} shows
a single wavelength $\hat\a_{\ell,k_1}^0$, simulated twice with an explicit, non-staggered Lax-Friedrich scheme. First, the initial perturbation is applied only to the phase fraction, \ie, $\hat u_{\phaseindex,k}^0\equiv 0$. Expression \eqref{eq:num_exp:init_cond} is used for the phase velocities in the second simulation. 
%An initial disturbance is seen in the simulation with the simpler initial conditions resulting from interactions between the two waves. 
%The quick wave is then seen to quickly dominate the wave growth.
After a short transition period where both waves interact, the fast wave is seen to dominate the growth of the first simulation. 
No transition period is observed in the second simulation.
Indeed, Liao et al.\ \cite{Liao_von_Neumann} demonstrated that the wave growth of a simulation with fairly random initial conditions quickly turns independent of these and develops according the the dominant wavelength. 
A figure similar to \ref{fig:num_exp:init_cond} is also presented in \cite{Brook_roll_wave_Godunov_comp_dressler} for gravity driven flows in collapsible tubes.
%
%Brook et al.\ demonstrate in \cite{Brook_roll_wave_Godunov_comp_dressler} that 
%%also flows whose initial conditions are not as carefully thought through
%also flows whose single wavelength initial conditions composes both waves quickly adjust to be governed by the dominating wave. 
%The effect of not choosing the initial conditions as carefully as described above is a disturbance in the growth trend during only the initial stage of the simulation. 
%

%https://se.mathworks.com/matlabcentral/newsreader/view_thread/240264?requestedDomain=www.mathworks.com

%\begin{figure}[h!ptb]%
%\centering
%%
%\begin{subfigure}{\columnwidth}
%\includegraphics[width=\columnwidth]{./images/growth_ylog_LF_UW_Roe_exp_imp.pdf}%
%\caption{}%
%\label{fig:num_exp:wave_growth}%
%\end{subfigure}
%%
%\begin{subfigure}{.8\columnwidth}
%\includegraphics[width=\columnwidth]{./images/Creal_LF_UW_Roe_exp_imp.pdf}%
%\caption{}%
%\label{fig:num_exp:wave_speed}%
%\end{subfigure}
%%
%\caption{}%
%\label{fig:num_exp}%
%\end{figure}

%\subsection{Closure Model}
\subsection{Test Case}
\label{sec:test_case}
The setup for the computational examples will now be presented. 
This setup is chosen fairly arbitrarily and corresponds to the experimental and numerical setup used in \cite{Holmaas_roll_wave_model, George_PhD,Akselsen_char_TEMP,Akselsen_roll_wave_TEMP}.
The friction closures $\tau\k$ and $\tau\_i$ are from the Biberg friction model as presented in \cite{Biberg_friction_duct_2007}, also described in the other citations just mentioned.
Fixed parameters are presented in Table~\ref{tab:parameters} unless otherwise stated
and constitutes a high-pressure, positively inclined flow.
The flow state is chosen such that the flow is weakly unstable according the differential VKH criterion. 
Truly fixed flow parameters in the incompressible flow simulations are the steady state level height $\overline h = 0.2 d$ and mixture velocity $\Qm/\area = 3.4 \unitfrac ms$.
The equivalent mean liquid area fraction is $\Al/\area = 0.142$ and
the chosen friction closures will yield the steady state superficial velocities $\USL \approx \unitfrac[0.154]ms$ and $\USG \approx \unitfrac[3.245]ms$.
The overall properties of the friction closure does not affect the linear stability analysis; only their resulting steady state $\S=0$ and its state derivatives enter into it.
In this particular case we have
% $\SAl=\unit[1.39\Ep6]{N\,m^{\text -5}}$, $\SQl\unit[=-9.07\Ep5]{N\,s}$ and $\SQg=\unit[7.63\Ep4]{N\,s}$.
%$\SAl=\unit[1.39\Ep6]{N/m^5}$, $\SQl\unit[=-9.07\Ep5]{N\,s}$ and $\SQg=\unit[7.63\Ep4]{N\,s}$.
 $\SAl=\unitfrac[1.39\Ep6]{kg}{m^6 s^2}$, $\SQl\unitfrac[=-9.07\Ep5]{kg}{ms}$ and $\SQg=\unitfrac[7.63\Ep4]{kg}{ms}$.

Only numerical parameters are varied in the tests provided in this section. 
The wavelength is therefore fixed at 30 diameters and the cell lengths are varied. 
Of course, doing it the other way around would also be insightful, showing which wavelengths one can expect to see on any given grid arrangement.

\begin{table}[h!ptb]
\centering
\begin{tabular}{|rl|rl|}
\hline
liquid density&				$\rho\l$	&998&			$\unitfrac[]{kg}{m^3}$\\
gas density&				$\rho\g$	& 50&			$\unitfrac[]{kg}{m^3}$\\
liquid dynamic viscosity&	$\mu\l$ 	& 1.00\Em3&		$\unit{Pa\:s}$\\ %\unitfrac[]{kg}{ms}
gas dynamic viscosity&		$\mu\g$ 	& 1.61\Em5&		$\unit{Pa\:s}$ \\
internal pipe diameter&		$d$ 		& 0.1&			$\unit[]m$	\\
wall roughness&							&2\Em5& 		$\unit[]m$ \\
pipe inclination& 			$\theta$	& 1\textdegree&		$-$\\
mean level height&			$\ol h$	    & 0.02&			$\unit[]m$\\
mixture velocity& 			$\Qm/\area$	& 3.4&			$\unitfrac ms$\\
wavelength&					$\lambda$	& 3	&			$\unit m$\\
\hline
\end{tabular}
\caption{Fixed parameters.}
\label{tab:parameters}
\end{table}

The time steps are regulated using a \CFL{} number, which makes the time step length proportional to the grid cell length (see Remark~\ref{rem:convergence}.)
The spectral radius will typically be used for selecting the time step length in schemes where the characteristic information is computed, and the \CFL{} number chosen close to unity;
$\dt = \CFL{}\, \dx/\max_{j,\pm}|\lambda_j^\pm|$ with $\CFL{}=0.95$ has been adopted in the presented Roe and local Lax-Friedrich schemes.
%the presented Roe and local Lax-Friedrich schemes adopt $\dt = \CFL{}\, \dx/\max_{j,\pm}|\lambda_j^\pm|$ with $\CFL{}=0.95$.
One of the phase velocities is commonly used in schemes not based on the model eigenstructure. 
Following \cite{Liao_von_Neumann}, the liquid velocity is chosen to limit the time step for the staggered upwind and Lax-Friedrich schemes, $\dt = \CFL{}\, \dx/\max_{j}|\ulj|$, with $\CFL{}=0.5$.
Implicit scheme simulations are performed by iterating on the new state with a 0.5 relaxation factor. % during each time step.

\subsection{Predictions}
\label{sec:results}
\begin{figure}[h!ptb]%
\centering
\begin{subfigure}{\columnwidth}
\caption{Wave growth%
\textcolor{\markercolor}{: Perturbation wave amplitude  $\max_j \{\tilde\a\lj\n\}/\area$ vs.\ time.}}%
\includegraphics[width=\columnwidth]{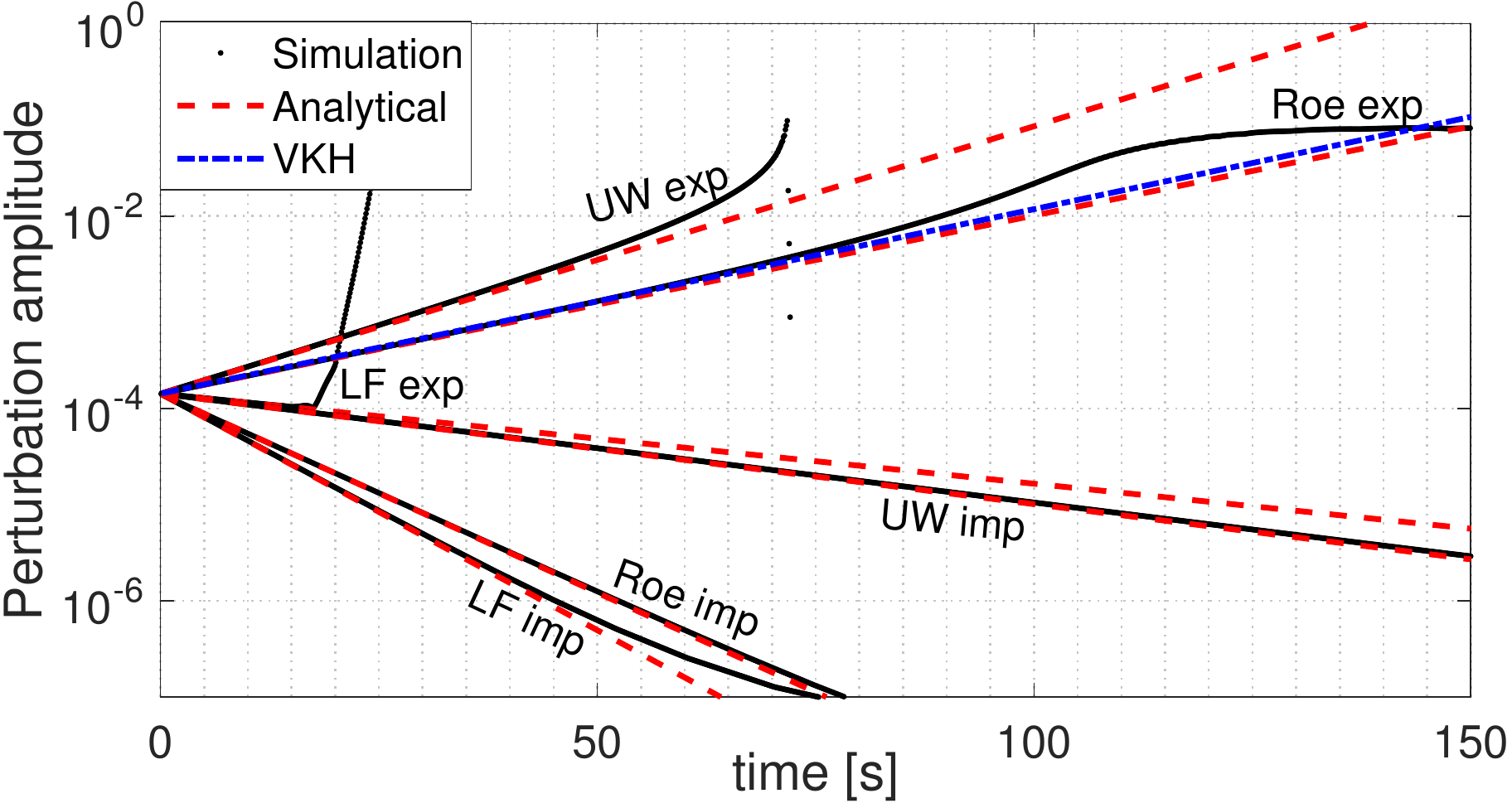}%
\label{fig:num_exp:wave_growth}%
\end{subfigure}
\begin{subfigure}{.9\columnwidth}
\caption{Wave celerity%
\textcolor{\markercolor}{: Perturbation peak location vs.\ time.}\\}%
\includegraphics[width=\columnwidth]{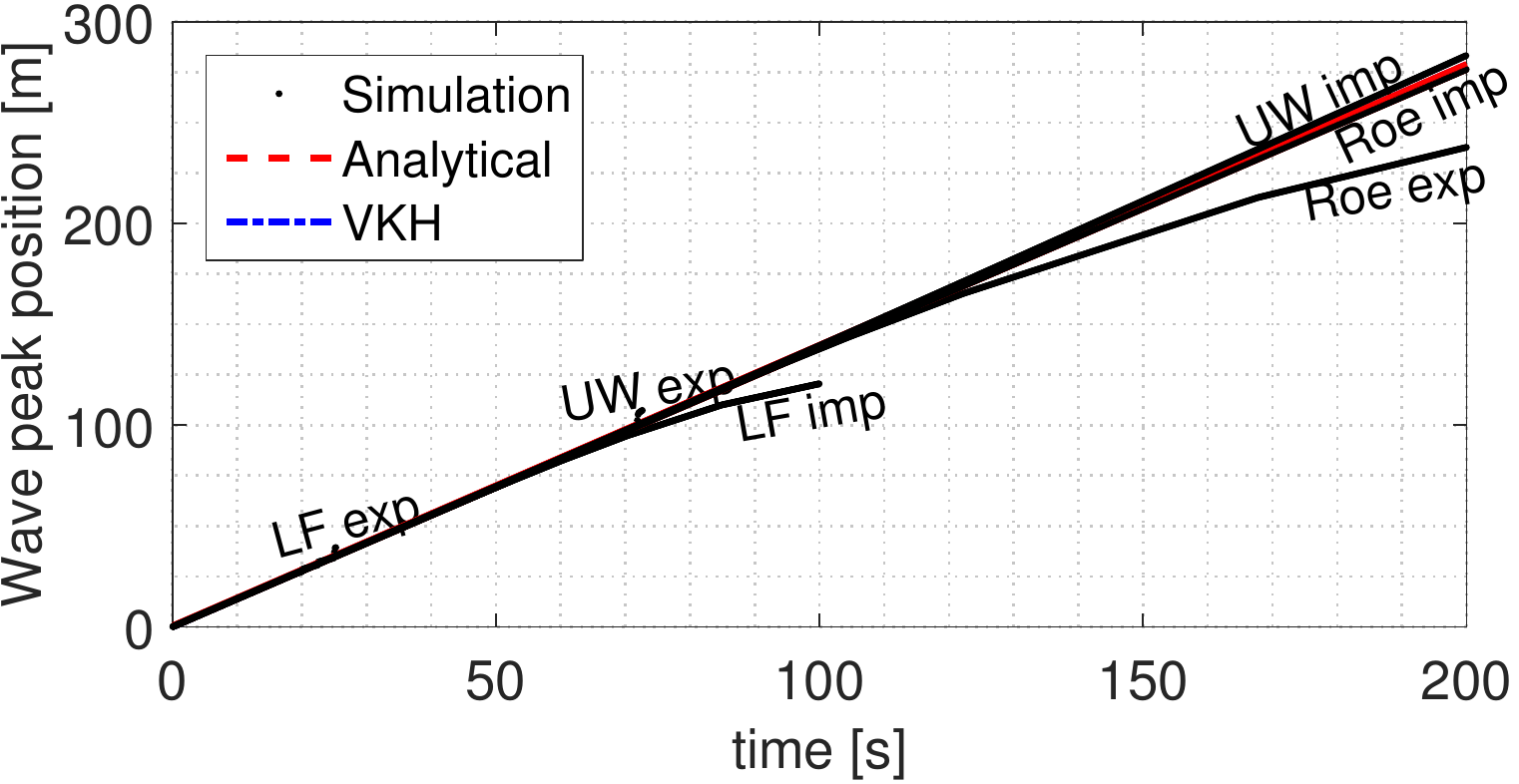}%
\label{fig:num_exp:wave_speed}%
\end{subfigure}
\caption{Numerical simulations vs.\ linear theory. Wavelength 30 $d$. 128 cells; $\phix = \pi/64$.}%
\label{fig:num_exp}%
\end{figure}

First, Figure~\ref{fig:num_exp} validates that the theory corresponds precisely to the linear growth of the discrete representations and shows the further development into the non-linear range.
The simulation domain here consists of 128 cells containing a single wave of the prescribed 30 diameter wavelength.
Subfigure~(\subref{fig:num_exp:wave_growth}) show the wave growth by logarithmically plotting 
the largest liquid fraction amplitudes. %  $\max_j|\wt \a_{\ell,j}\n|$.
%the highest liquid fraction in the domain minus the steady state fraction. 
The spatial locations of the wave crest peeks are plotted in Subfigure~(\subref{fig:num_exp:wave_speed}), providing the wave speed. 
Initial conditions are set to accommodate the most unstable wave, which is in these cases the fast wave.

\begin{figure}[h!ptb]%
\centering
\includegraphics[width=.9\columnwidth]{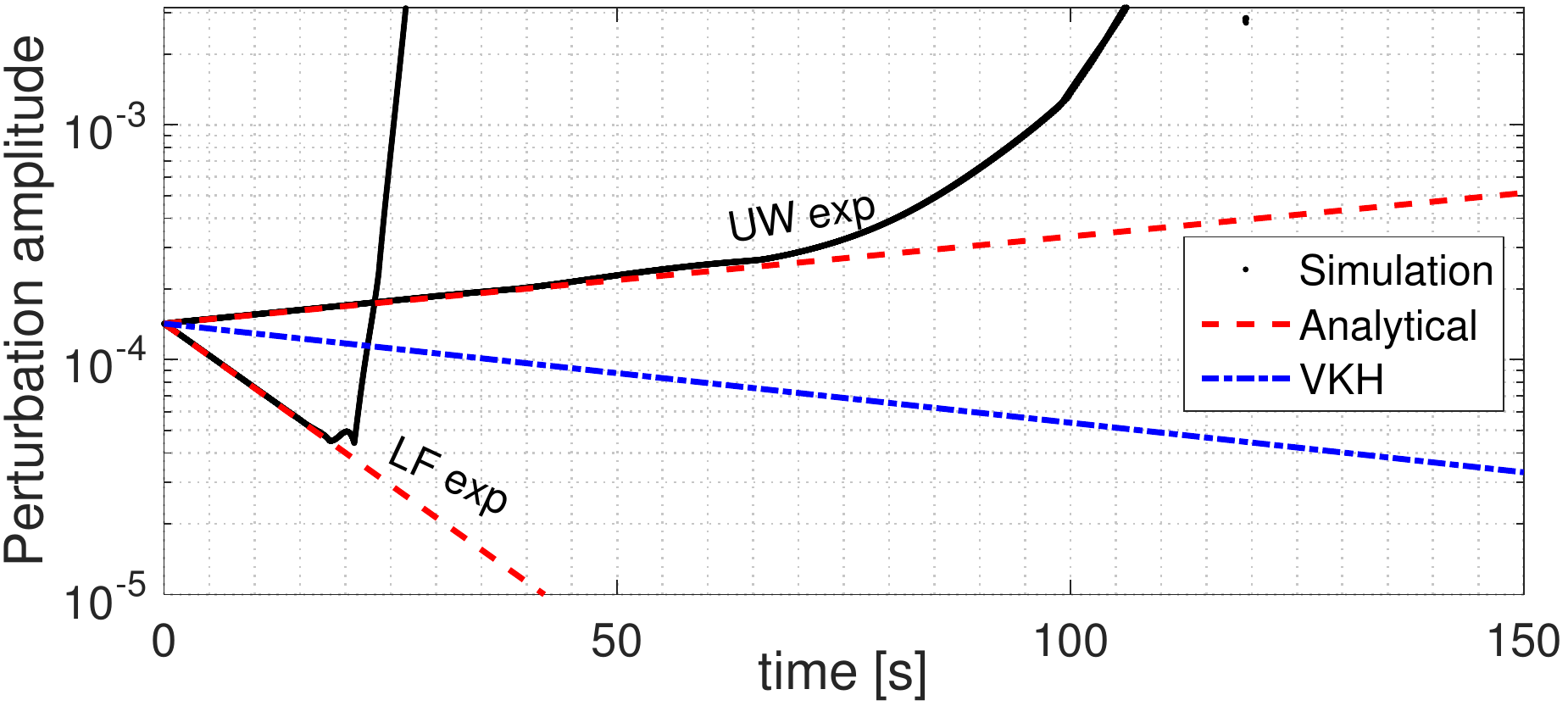}%
\\
\includegraphics[width=1\columnwidth]{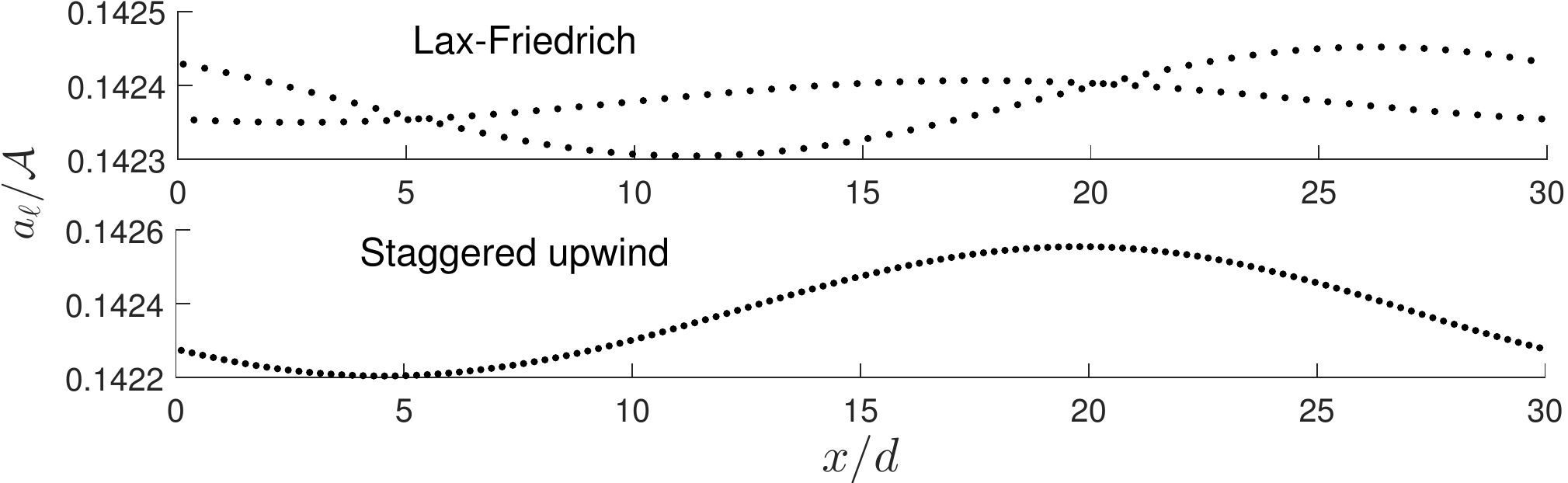}%

\caption{ Wave growth and accompanying snapshots from unstable Lax-Friedrich and staggered upwind simulations. The mixture velocity $\Qm/\area$ equals here $\unitfrac[3.1]ms$, at which the differential model is stable.}%
\label{fig:num_exp:LF_UW_instab}%
\end{figure}

The presented schemes provide a range of different behaviours.
We note immediately from Figure~\ref{fig:num_exp:wave_growth} that all implicit schemes are significantly more diffusive than their explicit counterparts, with numerical diffusion dominating the weak wave growth present in the differential solution. 
%All the implicit schemes are stable. 
Explicit versions of both the Lax-Friedrich scheme and the staggered upwind scheme eventually reach a final unstable state in which waves grow until the model is no longer hyperbolic and simulations crash.
% The  simulations crash shortly afterwards. 
They do so, however, in quite different manners. 
Where the Lax-Friedrich scheme first appears diffusive for then to be dominated by a high-wavenumber instability, the upwind scheme simply overpredicts the growth rate of the principle wave. 
This is further illustrated in Figure~\ref{fig:num_exp:LF_UW_instab}, showing growth rates and snapshots of simulations in which the differential model is stable due to a lower mixture velocity.
The former instability will usually be regarded as a `numerical instability,' commonly identified  by the sudden unphysical high-wavenumber growth. 
Determining, from visual inspection, whether the latter instability is `physical or not' is however not as straight forward as there are essentially no differences between the natural wave growth and the growth here attributed to numerical errors.

Lastly, the explicit Roe scheme is in Figure~\ref{fig:num_exp:wave_growth} seen to accurately match the continuous growth rate of the differential model. 
It also developers into a steady roll-wave solution, which is a valid solution for the differential problem (see \cite{Akselsen_char_TEMP};) 
Roe schemes are designed to be well adopted for strongly non-linear flows.
%That the Roe scheme remains stable also in the strongly non-linear range is not surprising as the scheme is designed to represent discontinuities.

The dispersion error of a 128 cell wave is very small, as seen in Figure~\ref{fig:num_exp:wave_speed}; wave crest positions are overlapping for as long as the waves remain in the linear range.
\\

%\begin{figure}[h!ptb]%
%\centering
%%
%\begin{subfigure}{\columnwidth}
%\caption{Wave growth}%
%\includegraphics[width=\columnwidth]{./images/growth_rollwave_LF_UW_Roe_exp_imp_Nj64.pdf}%
%\label{fig:num_exp:wave_growth}%
%\end{subfigure}
%%
%\begin{subfigure}{.9\columnwidth}
%\caption{Wave celerity}%
%\includegraphics[width=\columnwidth]{./images/Creal_LF_UW_Roe_exp_imp_Nj64.pdf}%
%\label{fig:num_exp:wave_speed}%
%\end{subfigure}
%%
%\caption{Numerical simulations vs.\ linear theory. Wavelength 30 $d$. 64 cells; $\phix = \pi/32$.}%
%\label{fig:num_exp}%
%\end{figure}

%\newpage

%\begin{figure}[h!ptb]%
%\includegraphics[width=\columnwidth]{./images/growth_rollwave_LF_UW_Roe_exp_imp.pdf}%
%\caption{Roll-wave case with Biberg. Wavelength 30 $d$. 128 cells; $\phix = \pi/128$.}%
%\label{fig:num_exp:wave_growth_roll_wave}%
%\end{figure}

%\section{Comparisons}

\begin{figure}[h!ptb]%
%\centering
%
\begin{subfigure}{\columnwidth}
\caption[t]{Wave growth $k\Im(\cdisc)-[\unitfrac 1s]$ }%
\includegraphics[width=\columnwidth]{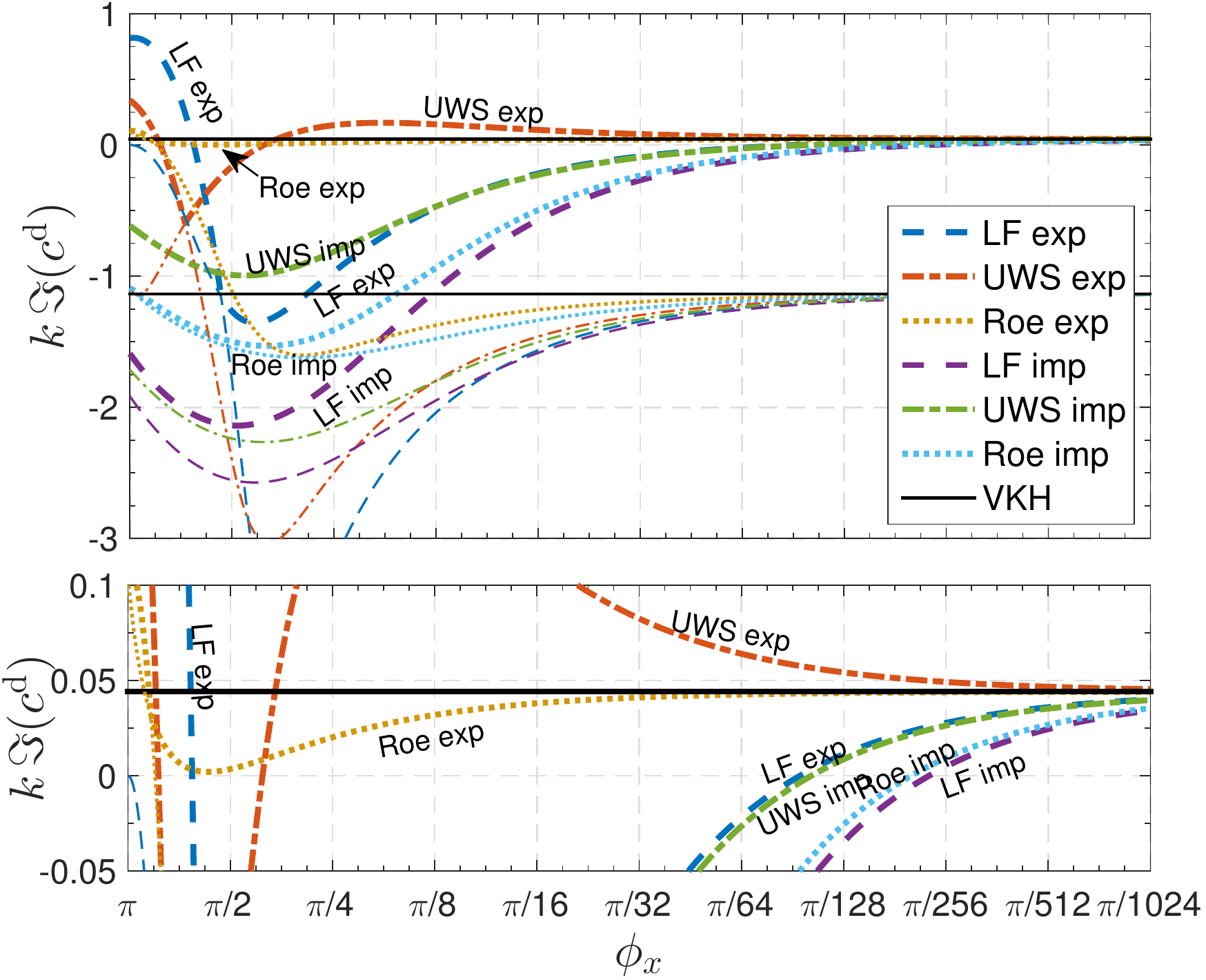}%growth_phi_rollwave_wavelength3_bothwave_theta0p017_USL0p15_USG3p2.pdf}%
\label{fig:comp:phi:wave_growth}%
\end{subfigure}
\begin{subfigure}{\columnwidth}
\caption[b]{Wave celerity $\Re(\cdisc)-[\unitfrac ms]$}%
\includegraphics[width=\columnwidth]{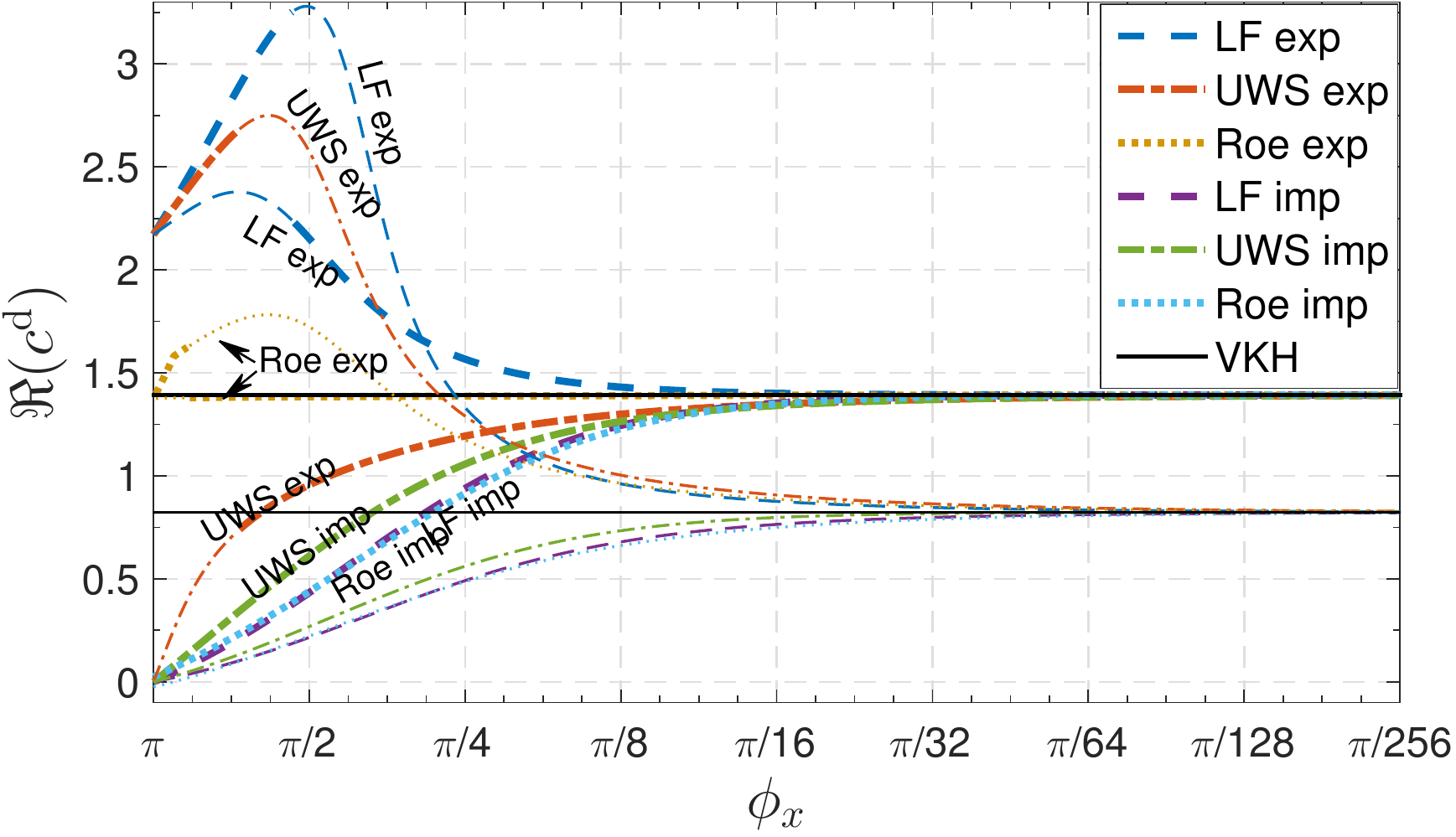}%
\label{fig:comp:phi:wave_speed}%
\end{subfigure}
\caption{Linear theory with varying cell lengths $\dx$. 30 $d$ wavelength. $\phix = k\,\dx = 2\pi/(\text{\# cells in a wave})$}%
\label{fig:comp:phi}%
\end{figure}

\begin{figure}[h!ptb]%
\centering
\includegraphics[width=.8\columnwidth]{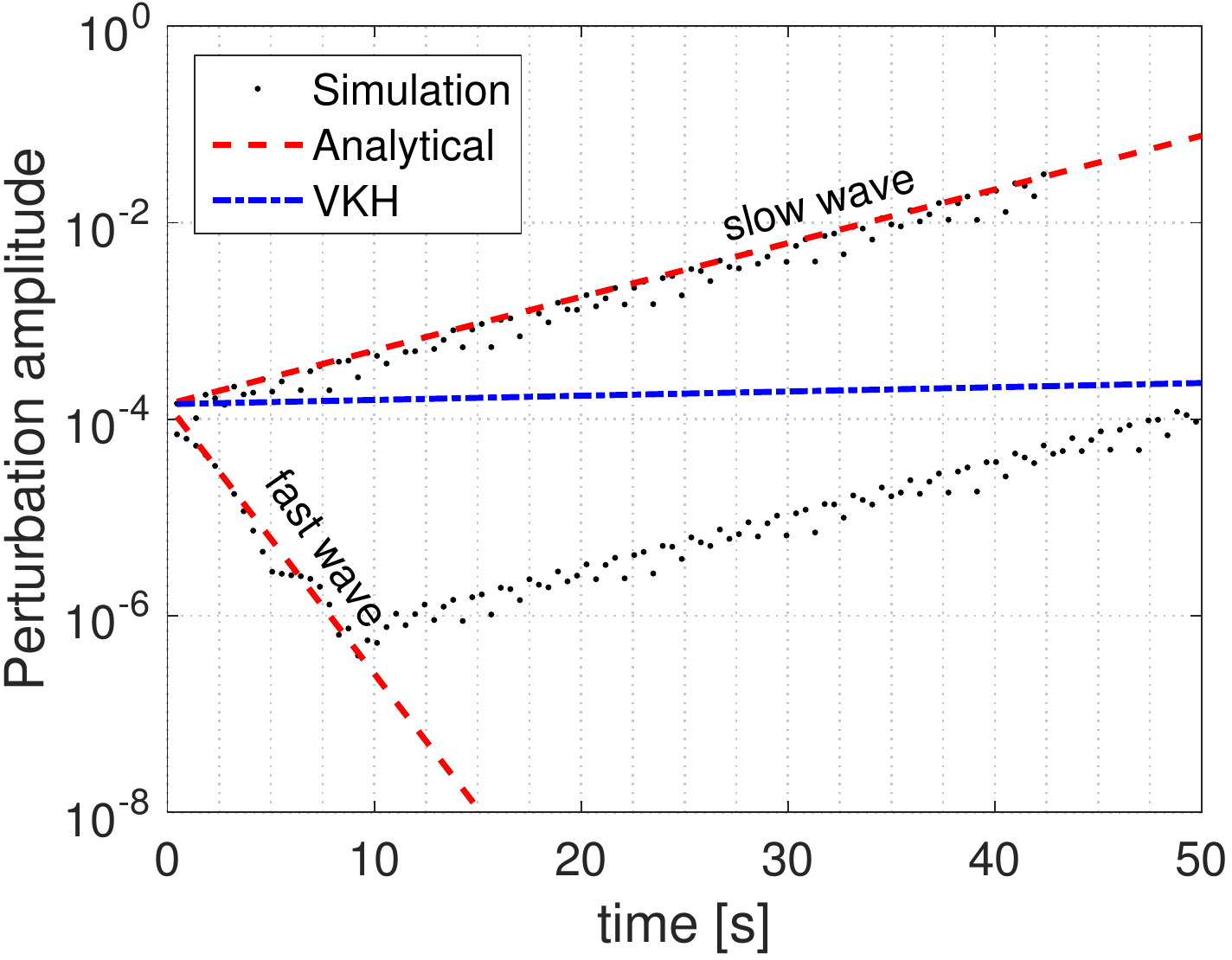}%
\caption{Explicit Lax-Friedrich scheme. Three cells per wave; $\phix = \frac23\pi$.}%
\label{fig:comp:LF_unstab}%
\end{figure}

Next we examine the response of the 30 diameter wave to changes in the spatial resolution. 
Figure~\ref{fig:comp:phi} shows the wave growth, or \textit{pulsation}, $k\, \Im(\cdisc)\,[\unitfrac1s]$ and the wave celerity $\Re(\cdisc)\,[\unitfrac ms]$.
Both the fast and the slow waves are shown, a bold line used for the wave with the higher growth rate.
Subfigure~(\subref{fig:comp:phi:wave_growth}) shows that the Roe scheme gives very accurate growth results for $\phi \leq \pi/32$ (using 64 cells or more)
and predicts wave growth for all $\phi$. 
An accurate celerity for the fast wave is observed in Subfigure~(\subref{fig:comp:phi:wave_speed}) for all $\phi$.
The explicit Lax-Friedrich and implicit staggered upwind schemes start predicting wave growth 
%between $\phi = \pi/64$ and $\phi = \pi/128$
around $\phi = \pi/100$, and the implicit Roe and implicit Lax-Friedrich schemes start doing so
% between $\phi = \pi/128$ and $\phi = \pi/256$. 
around  $\phi = \pi/210$.
The explicit upwind scheme overpredicts the wave growth everywhere above $\phi=\pi/3$.
Finally, we see that the explicit Lax-Friedrich scheme becomes unstable around $\phi\gtrsim\tfrac23\pi$, and that it is then the slow wave that has taken over the growth. 
In fact, from Subfigure~(\subref{fig:comp:phi:wave_speed}) we see that the `slow' wave moves faster here. %, though it may not make as much sense talking about the wave speed direction here at the wave only consists of three cells and less.
This very particular finding is tested and confirmed in Figure~\ref{fig:comp:LF_unstab}, where three-celled waves ($\phix=\frac23\pi$) are applied as the initial condition. 
Two simulations are shown wherein the initial velocity condition \eqref{eq:num_exp:init_cond} is made to accommodate the fast wave in one and the slow wave in the other.
The simulations follow the predicted growth behaviour precisely, with the slow wave growing and the fast wave diminishing. 
After a while though, also the fast wave simulation becomes unstable as the slow wave grows from out of numerical inaccuracies.
The plotted points are the maximum amplitudes of the spatial node sets; there is some scattering of these points as the simulated waves are represented by a regularly alternating three-point pattern.
\\

\begin{figure}[h!ptb]%
\centering
\begin{subfigure}{\columnwidth}
\caption{Wave growth $k\Im(\cdisc)-[\unitfrac 1s]$}%
\includegraphics[width=\columnwidth]{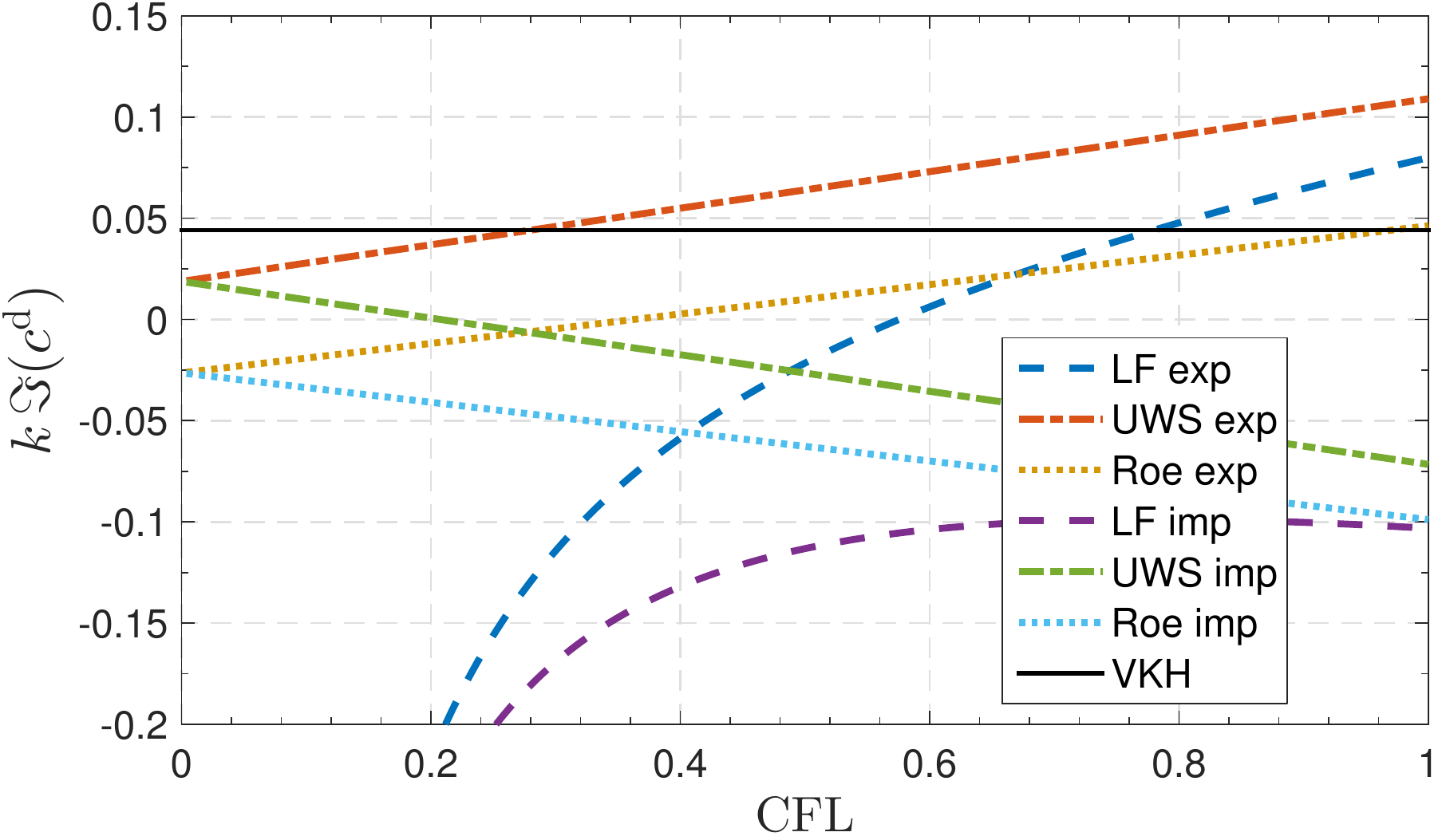}%
\label{fig:comp:CFL:wave_growth}%
\end{subfigure}
\begin{subfigure}{\columnwidth}
\caption{Wave celerity $\Re(\cdisc)-[\unitfrac ms]$}%
\includegraphics[width=\columnwidth]{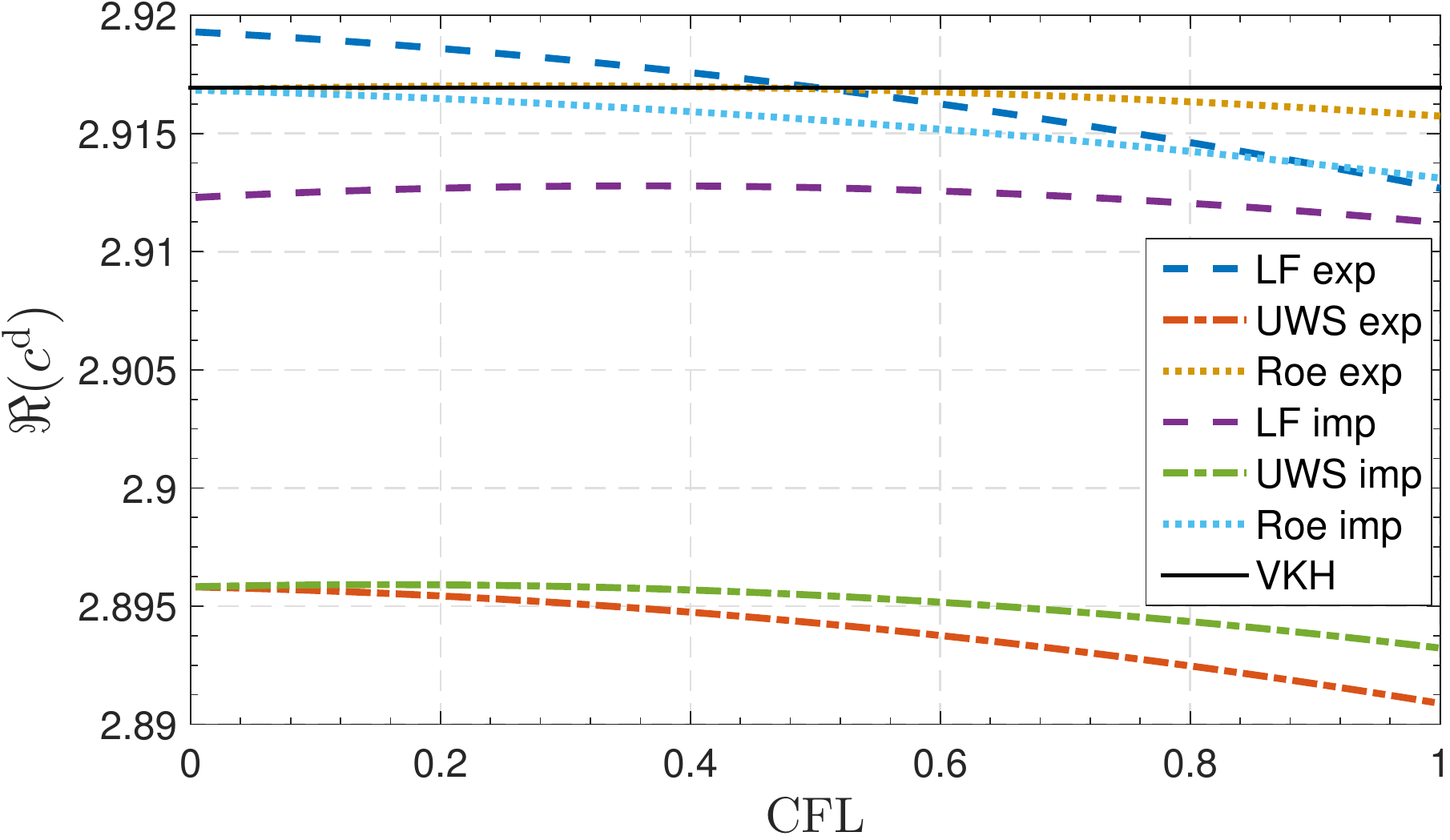}%
\label{fig:comp:CFL:wave_speed}%
\end{subfigure}
\caption{Linear theory with varying \CFL{} number, showing the fast wave. 128 celled wave; $\phix = \pi/64$}%
\label{fig:comp:CFL}%
\end{figure}

Next we look at how the discrete representations respond to changes in the time step. 
We have already observed that there is a significant difference in the stability and diffusivity of explicit and implicit time step integration. 
A strong time step dependence was also noted in \cite{Akselsen_char_TEMP} for both linear and non-linear wave simulations.
Figure~\ref{fig:comp:CFL} shows the wave growth and wave celerity as functions of the \CFL{} number. 
Time steps $\dt$ are selected on the basis of these as per the individual method descriptions, 
\ie, based on the spectral radius $\max |\lambda^\pm|$ in the Roe schemes and on the liquid velocity $u\l$ in the Lax-Friedrich and upwind schemes.

\begin{figure}[h!ptb]%
\centering
\includegraphics[width=\columnwidth]{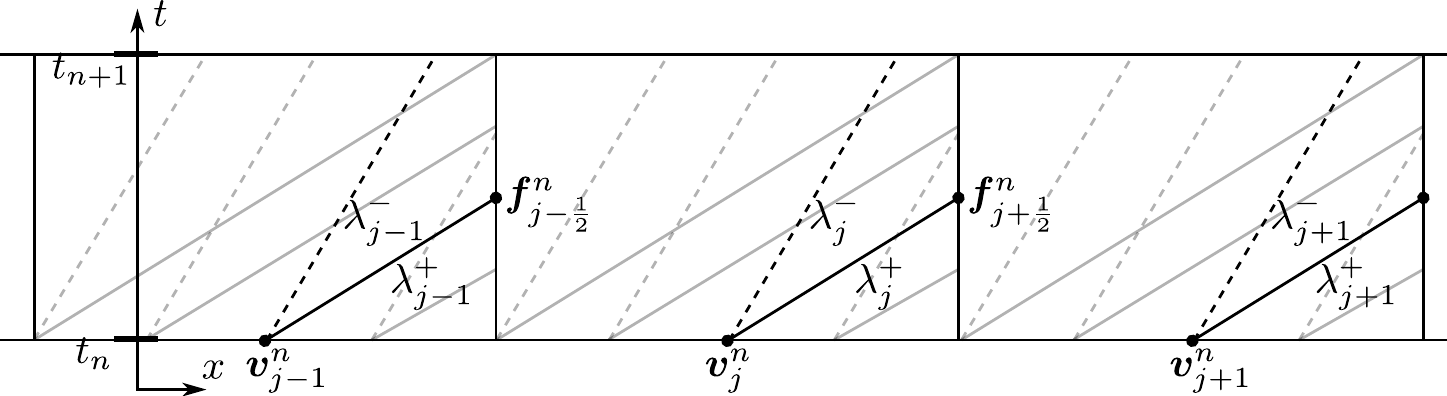}%
\caption{Sketch of the characteristic paths at $\CFL{} = 1$}%
\label{fig:comp:CFL_1}%
\end{figure}

We first note that the explicit and implicit versions of  all schemes converge towards the same growth and celerity as the \CFL{} number approaches zero, except for the Lax-Friedrich scheme whose numerical diffusion is inversely proportional to the time step length and thus approaches infinity with reducing \CFL{} number.
Growth increases `with increasing explicitness' for the Roe and upwind scheme.
%, though not for the Fax-Friedrich scheme as the numerical viscosity is inversely proportional to the time step length. 
%\{As noted in \cite{Akselsen_char_TEMP}\}, growth predictions seems to be accurate in explicit schemes based on characteristic information with a \CFL{} number near unity, in this case the Roe scheme.
As noted in \cite{Akselsen_char_TEMP}, growth predictions seems to be accurate in explicit schemes based on characteristic information, in this case the Roe scheme, as the \CFL{} number nears unity.
%It is quite possible that the same can be said about schemes which are not based on characteristic information, and that the spectral radius can be used to estimate an optimal time step also in non-characteristic schemes. 
\textcolor{\markercolor}{
An interpretation of why this is is shown in Figure~\ref{fig:comp:CFL_1};
information from the upwind mean state would spread nicely
over the cell face during the time step limited by $\CFL{} \approx 1$.
This is particularly so if the cell face fluxes are dominated by information travelling along the path of the quickest characteristic, which seem to be case for supercritical flows.
}

The celerity graphs shown in Figure~\ref{fig:comp:CFL:wave_speed} show that the wave celerity of the Roe scheme becomes increasingly accurate as the time step is reduced.
%Indeed, the Roe scheme is designed to provide local solutions of the linearized Riemann problem and it obeys the Rankine-Hugoniot shock condition.
Indeed, the Roe scheme is designed to provide the accurate shock speeds for non-linear problems.
Both centred schemes (Roe and Lax-Friedrich) provide better estimates of the wave celerity than the upwind scheme, whose dispersion error is mostly caused by the $\eEuler^{-\imunit \phix/2}$ term generated by the spatially asymmetric upwind formulation.

\begin{figure}[h!ptb]%
\centering
\includegraphics[width=.9\columnwidth]{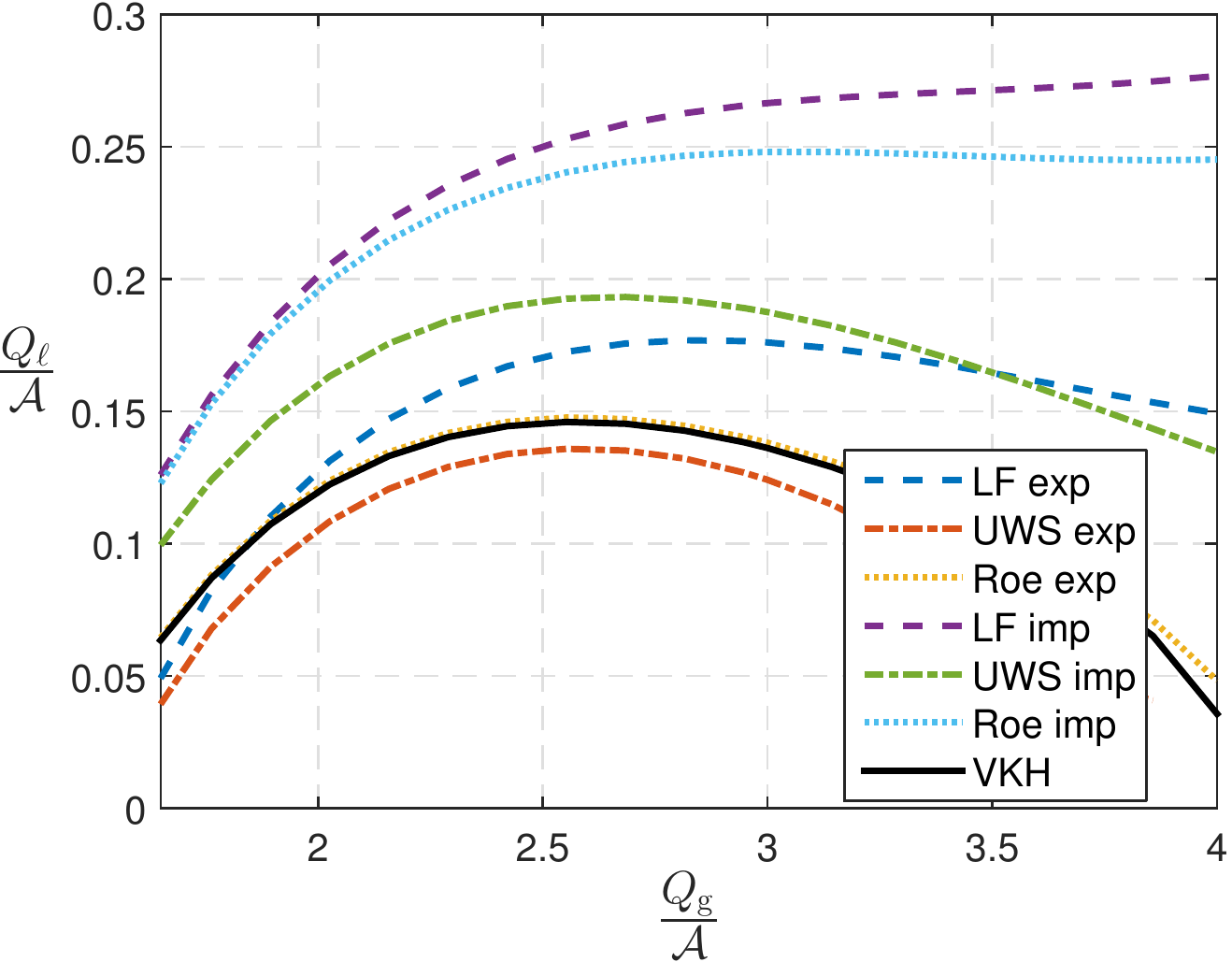}%
\caption{Flow map predicted by various schemes. $\phix = \pi/64$.}%
\label{fig:flow_map_disc}%
\end{figure}

An alternative way of presenting the schemes' ability to predict wave growth is through a flow map such as that presented in Figure~\ref{fig:flow_map_disc}.
%The flow rates are now varied with the wavelength, cell length and \CFL{}-number kept constant
This map is of the same 30 diameter wave with 128 cells ($\phix = \pi/64$,) and the flow rates are now varying. 
The \CFL{}-numbers are specified as before and a root searcher is employed to identify the critical point. 
Numerically searching for the critical state makes this form of visualization more computationally costly than the other plots presented in this section, which were all explicitly computed.
Note that Equation~\eqref{eq:VKH_crit} is not valid in the discrete representations because $\cnudisc$ does not equal $\cdisc$.
A consequence is that the marginal stability of a discrete representation is wavelength dependent, while the marginal stability of the differential model is not.
\\

The local Lax-Friedrich scheme, not included amongst the presented results,
gave growth results nearly identical to that of the Roe scheme all over. 
This indicates that it is the characteristic information in the viscous term and time step, rather than the off-diagonal contribution, that accounts for the favourable growth results of the Roe scheme.
The local Lax-Friedrich scheme does however not converge as precisely with respect to the wave celerity $\Re(\cdisc)$ as the Roe scheme, slightly underpredicting the fast celerity in plots similar to those of Figure~\ref{fig:comp:phi:wave_speed} and \ref{fig:comp:CFL:wave_speed}. This scheme was also seen to the have the same type of slow wave instability near $\phi=\pi$ as the simple Lax-Friedrich scheme.
\\

\begin{figure}[h!ptb]%
\centering
\begin{subfigure}{\columnwidth}
\caption{Wave growth $k\Im(\cdisc)-[\unitfrac 1s]$}%
\includegraphics[width=\columnwidth]{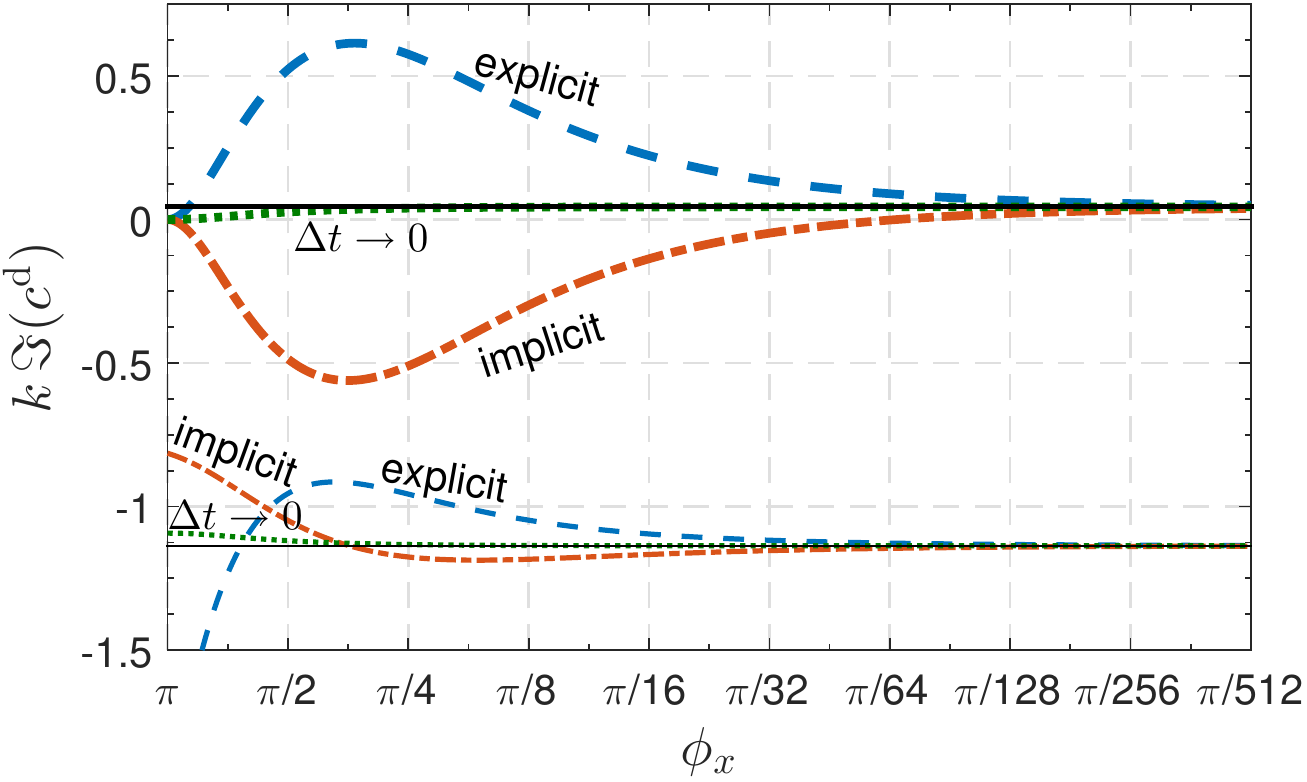}%
%\\\includegraphics[width=\columnwidth]{./images/growth_phi_rollwave_wavelength3_CD_NS_exp_vs_imp_criticalpoint.pdf}
\label{fig:comp:centered:wave_growth}%
\end{subfigure}
\begin{subfigure}{\columnwidth}
\caption{Wave celerity $\Re(\cdisc)-[\unitfrac ms]$}%
\includegraphics[width=\columnwidth]{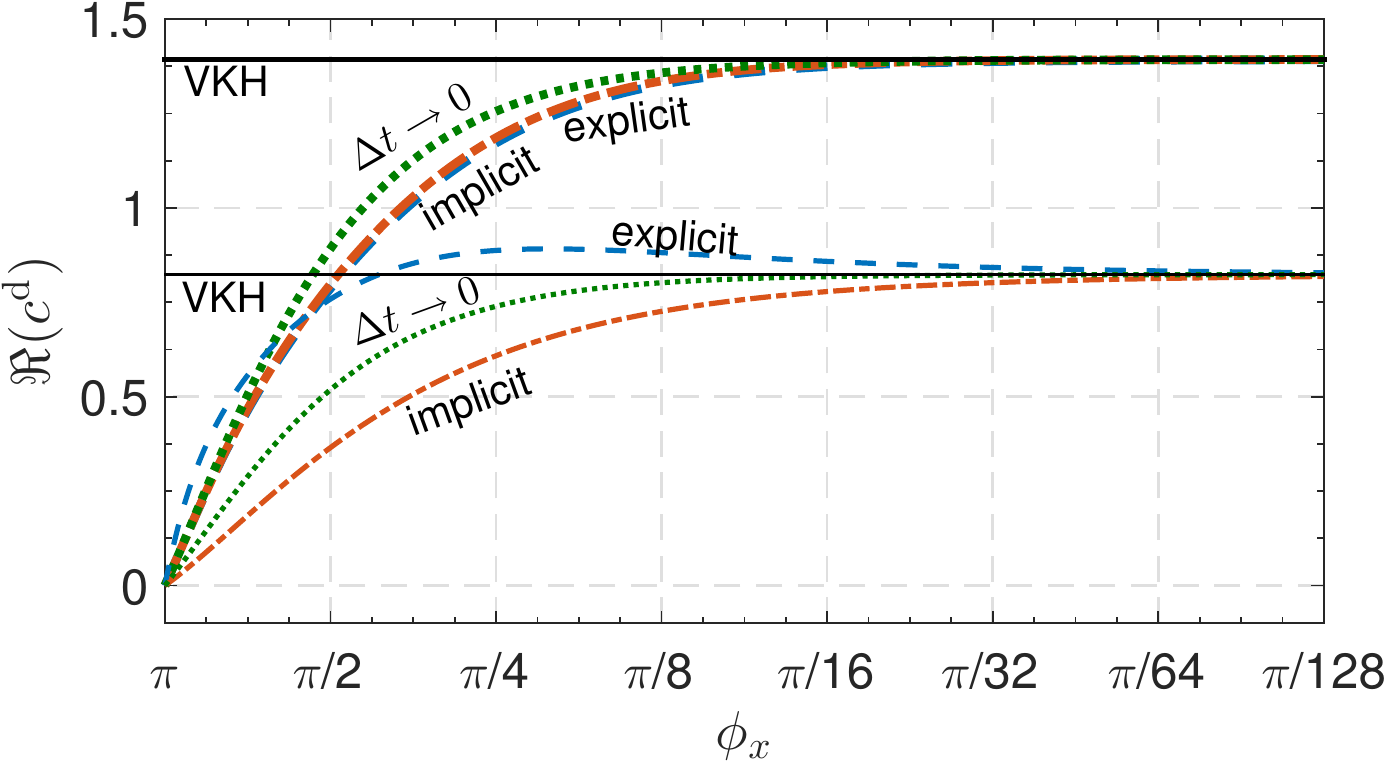}%
%\\\includegraphics[width=\columnwidth]{./images/speed_phi_rollwave_wavelength3_CD_NS_exp_vs_imp_criticalpoint.pdf}
\label{fig:comp:centered:wave_speed}%
\end{subfigure}
\caption{Non-staggered centred difference scheme with $\nu=0$.}%
\label{fig:comp:centered}%
\end{figure}

%Finally we look a bit closer at the effect of the time discretization. 
So how does the time integration affect the stability and diffusivity of our solution?
%In the complex $c$ plane, such as that shown in Figure~\ref{fig:IKH_vs_VKH}, the difference between explicit/implicit discretizations is a convex/concave curving of all lines about the origin.
Let us examine the influence of the time discretization on a centred, non-staggered scheme without artificially added viscosity (using $\nu=0$ in the  Lax-Friedrich scheme.)
Such a scheme, though on a staggered grid, was in \cite{Liao_von_Neumann} deemed the most accurate amongst those tested.
%, although the way in which the time discretization played into this was not made very clear.  
Figure~\ref{fig:comp:centered} shows the growth rates and wave celerities of this centred scheme with an explicit and implicit time integration, the liquid velocity based CFL number equalling $0.5$. 
The curves for when the time step approaches zero are also shown. 
Note first that the high-wavenumber slow-wave instability of the explicit Lax-Friedrich scheme is not present here, attributing that phenomenon to the artificial viscosity $\nu$.
The growth rates of the fast wave in Subfigure~(\subref{fig:comp:centered:wave_growth}) are near mirror images of each other, the numerical error being predominantly attributed to the time step integration. 
This is important to be aware of; 
numerical diffusion errors are often thought of as a symptom of the \textit{spatial} discretisation alone.
%numerical diffusion errors are often thought of purely in consideration of the \textit{spatial} discretisation.
%numerical diffusion errors are often considered an issue purely associated with the \textit{spatial} discretisation.

If the wave growth in this example is dominated by the time integration then we would expect accurate growth results for small time steps $\dt$.
In fact, since $\ddisc_x=\tfrac \imunit \dx\sin\phix$ is purely imaginary, 
examining the dispersion equation \eqref{eq:disp_eq}-\eqref{eq:dS} reveals that 
%$\cnudisc\equiv c\_{crit}$ 
$\cnudisc$ equals the critical VKH celerity exactly 
at the state of marginal VKH stability.
%$\cnudisc\equiv c\_{crit}$ at the state of marginal VKH stability, as can be deduced form the dispersion equation \eqref{eq:disp_eq}-\eqref{eq:dS}. 
The discrete growth rates at marginal stability will then be the exact VKH growth if also $\ddisc_t$ is purely imaginary, which is the case if $\dt\rightarrow0$.
Wave growth shown in Subfigure~\ref{fig:comp:centered:wave_growth} for vanishing time steps is thus very close to the differential growth as the considered state is close to the marginally stable state.
%The state simulated is very close to the critical VKH condition so that the growth of the $\dt\rightarrow0$ case is very close to the VKH growth. 
Spatial discretisation errors are manifested foremost in the real components of $c$, \ie, the wave speeds, which are often deemed to be of secondary importance.
\\

%Let us further consider the the stability errors associated with the time integration.
%We may generalize the discrete time differential operators listed in Table~\ref{tab:delta} by introducing
% the `degree of implicitness' $r$ as a linear combination of the forwards and backwards Euler integrations, 
%$\tfrac1\dt\br{\psi\nn-\psi\n}+r(\cdots)\nn+(1-r)(\cdots)\n$.
%The discrete differential operator for this integration is $\ddisc_t=\frac 1\dt\frac{\exp\br{-\imunit  k \cdisc}-1}{r\br{\exp\br{-\imunit  k \cdisc}-1}+1}$,
%giving $\cdisc = \tfrac1{k\dt} \ln\br{\frac{1+(1-r)\ddisc_t\dt}{1-r\ddisc_t\dt}}$.
%$r=0$ here constitutes the explicit time integration, $r=1$ the implicit one and $r = \frac 12$ the Crank-Nikolson time integration.
%Inserting the exact operator $\ddisc_t=\delta_t=-\imunit k c$ then gives the pure time integration error for varying degrees of implicitness.

\begin{figure}[h!ptb]%
\includegraphics[width=\columnwidth]{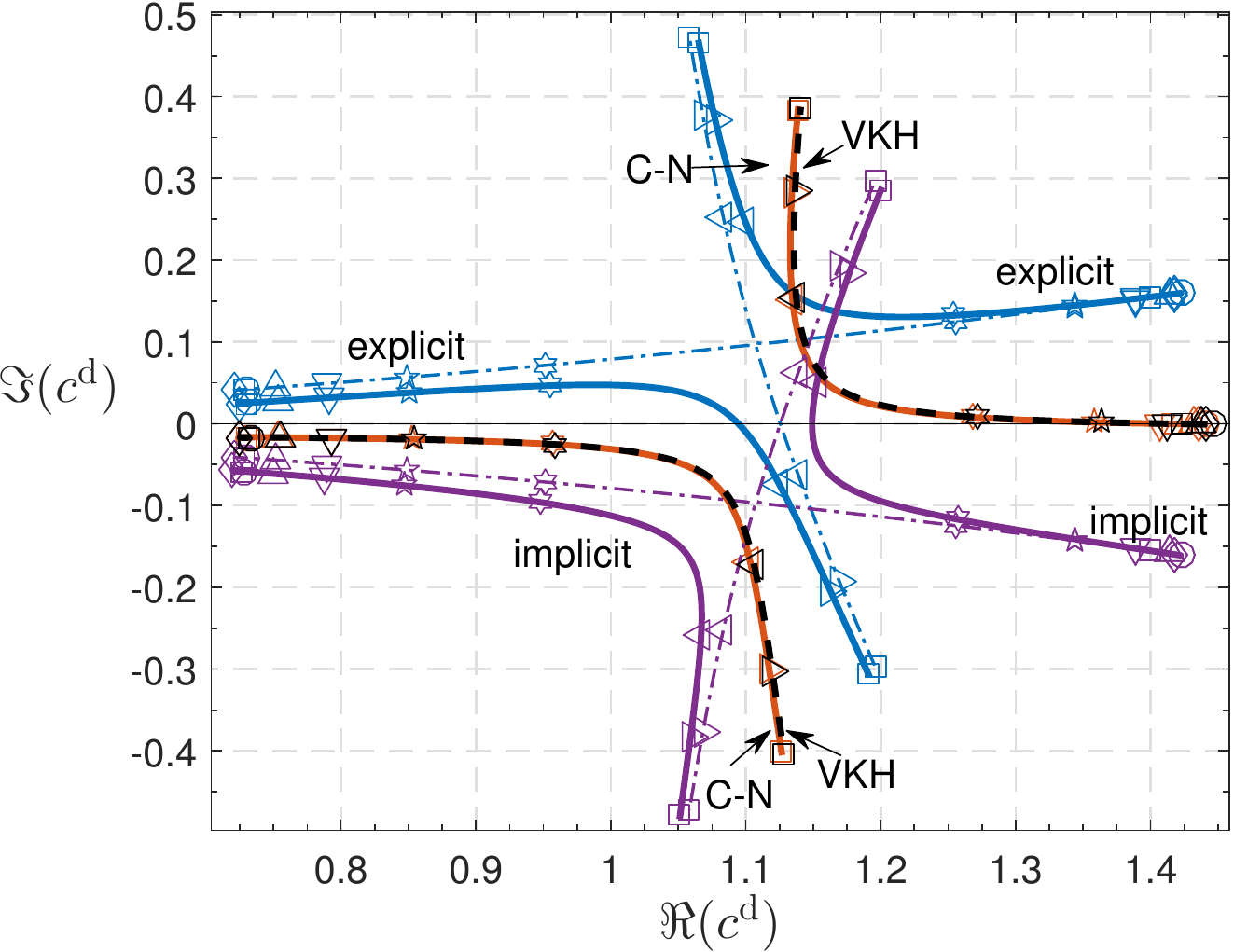}%
\caption{
Complex celerity $c$ under varying superficial gas velocity $\USG$, plotted with time integration error only ($\ddisc_t = \delta_t$.) 
Explicit, implicit and Crank-Nicolson time integrations are shown, cf.\ Table~\ref{tab:delta}, along with the differential VKH celerity.
Thinner lines (dot-dashed) show the corresponding explicit and implicit IKH celerities (where $S\equiv0$.)
The considered wave is one diameter long and $\dt = \unit[0.0025]s$.
State parameters and markers for the superficial gas velocities are the same as in Figure~\ref{fig:IKH_vs_VKH}
}%
\label{fig:comp:cplot_imp_vs_exp}%
\end{figure}

Figure~\ref{fig:comp:cplot_imp_vs_exp} shows plots similar to the celerity plots shown back in Section~\ref{sec:VKH}, Figure~\ref{fig:IKH_vs_VKH}, where  viscous and inviscid Kelvin-Helmholtz celerities were compared under varying gas rates.
The plots in Figure~\ref{fig:comp:cplot_imp_vs_exp} show the errors form the time integration only, \ie, $\ddisc_t=\delta_t=-\imunit k c$, equivalent to the error as $\dx$ approaches zero.
A somewhat shorter wave, one diameter in length, has here been chosen to highlight the time integration effect, and the time step is $\dt = \unit[0.0025]s$.

The general trend of the time integration error is most easily observed from the thinner lines in Figure~\ref{fig:comp:cplot_imp_vs_exp}, showing the celerities in the inviscid case $S\equiv0$. 
These lines would follow the abscissa if not for the error, as in Figure~\ref{fig:IKH_vs_VKH}.
The effect of the time integration error is to curve the celerity lines about the origin, convexly if the integration is foremost directed forwards in time (explicit,) and concavely if it is foremost directed backwards in time (implicit.) 
Errors are thus seen to increase with $\Re(\cdisc)$, which for the fast wave is means that it increases with decreasing $\USG$.
Crank-Nicolson integration, abbreviated `C-N' in the figure, is seen to be quite accurate in this case.

%\section{Discussion and Conclusions}
\section{Concluding Remarks}
\label{sec:conclusions}
Practically no computational effort is associated with the linear stability expressions.
The examples presented in the previous section demonstrate some of the information instantaneously available through use of linear theory. 
This theory was shown to give quantitative discretization requirements for obtaining physical wave growth for any given wavelength. 
Conversely, it tells us what wavelength we can expect to see grow on any given grid, and how these waves move and grow in the discrete representation compared to in the differential model. 
Numerical errors were shown to manifest in both suppressed and excited growth, the latter being related to what is usually referred to as \textit{numerical instabilities.} Some such instabilities were also observed in the low-wavenumber range where there is very little distinction between numerical and physical instabilities.
Discrete stability theory was further able to demonstrate the way in which the time discretization affects predicted stability and to indicate appropriate values for the \CFL{} number.

This information can aid in choosing reliable simulation parameters prior to simulation, and 
may give insight into whether or not simulation results can be considered physical.
%may prove useful in interpreting  simulation results and the degree to which they may be considered physical.

\bibliographystyle{plainnat}%{plain}% apsrev
{\footnotesize
\bibliography{refs_PhD}}

%\appendices

%\newpage
%\onecolumn
%\section{Discrete Dispersion Relations} 
%\label{sec:collected}
%\input{collected_disp}
%\newpage
%\input{collected_wave}
%\newpage
%\input{collected_stab}

%\section{Alternative schemes}
%\label{sec:alternative_schemes}
%\input{SLUGGIT_mass_reduction}
%\input{staggered_conservative_partial_UW}
 
%\input{Barnea_model}

\appendix

\section{A Roe Scheme}
\label{sec:Roe_scheme}
The Roe scheme is commonly written
\[
\frac{\bv_j\nn - \bv_j\n}\dt + \frac{\bf\jph-\bf\jmh}{\dx}  = {\bs_j}
\]
where $\bf\jpmh$ are the fluxes in the solution of the linearized Riemann problem 
\begin{equation}
\begin{gathered}
 \bv_t +  \Big(\Jac\Big)\^{Roe}\jph \bv_x = 0 %\hspace{-5mm}\of{\bv\_R,\bv\_L } 
\\
\bv\of{x,0} = \bv_j,\: (x<0);\quad \bv\of{x,0} = \bv\jp,\: (x>0)
\end{gathered}
\label{eq:Roe_problem}
\end{equation}
at each cell face $x\jph$.  
$\Jac\^{Roe}\jph$ is the Roe-average of the Jacobian \eqref{eq:base_model:Jac}, constant in each Riemann problem.
Using arithmetic mean notation $\ol w = \frac 12(w_j+w\jp)$,
the Roe matrix of model \eqref{eq:base_model} is 
\[
\Jac\^{Roe}\jph = \left.\pdiff\bf\bv\right|_{\ol\bw, \widetilde \HofAl'};%\br{\ol \a\l, \ol \a\g, \ol u\l,\ol u\g}
\]
the Jacobian evaluated with the mean primitive variables 
\[
\ol\bw = \br{\ol \a\l, \ol \a\g, \ol u\l,\ol u\g}^T
\]
and 
\[
\widetilde \HofAl' = 
\begin{cases}
\frac{h\jp-h_j}{\aljp-\alj}, & \aljp\neq\alj \\
\dHdAl\of{\ol\a\l},& \text{otherwise}
\end{cases}
\]
replacing $\dHdAl$.

The flux solution of \eqref{eq:Roe_problem} may be written 
\begin{equation}
\bf\jph 
%\bf\of{0,t}
= \ol \bf -\tfrac12\LL\inv |\Lamb| \LL (\bv\jp-\bv_j)
\label{eq:Roe_sol_f}
\end{equation}
where
\begin{align*}
|\Lamb| &= 
\begin{pmatrix}
	\big|\lambda^+\big| & 0 \\
	0 & \big|\lambda^-\big|
\end{pmatrix},
&
\text{and}&&
\LL &= 
\begin{pmatrix}
	1 & 1/\symkappa  \\
	1& -1/\symkappa
\end{pmatrix}
\end{align*}
are the absolute eigenvalue and eigenvector matrices of $\Big(\Jac\Big)\^{Roe}\jph$, respectively.
These are evaluated from \eqref{eq:eigenvalues} and \eqref{eq:kappa} at the mean state $\ol \bw$ with $\widetilde \HofAl'$ replacing $\dHdAl$.

\section{Transformation Matrices}
Conversion between primitive variables $\bw=\br{\al,\ag,u\l,u\g}^T$ and conservative variables $\bv=\br{\al,\diffk{\rho u}}^T$ of the \textit{incompressible} system (obeying \eqref{eq:sum_a_au}) is often useful. Transformation matrices for these are
\begin{subequations}
\begin{align}
\pdiff\bV\bW &= 
\begin{pmatrix}
	1 & 0 & 0 & 0 \\
	0 & 0 & \rho\l & -\rho\g
\end{pmatrix},%
\label{eq:transformation_matrices:dvdw}%
%\\
%\pdiff\bW\bV &= 
%\frac{1}{\Ag\rho\l + \Al\rho\g}\begin{pmatrix}
%	\Ag\rho\l + \Al\rho\g & 0 \\
%	-(\Ag\rho\l + \Al\rho\g) & 0 \\
%	\rho\l\br{U\g-U\l} & \Ag\\
%	\rho\g\br{U\g-U\l} & -\Al
%\end{pmatrix}.
\\
\pdiff\bW\bV &= 
\frac{1}{\Al\Ag\rho\m}\begin{pmatrix}
	\Al\Ag\rho\m & 0 \\
	-\Al\Ag\rho\m & 0 \\
	\rho\g\br{U\g-U\l} & \Ag\\
	\rho\l\br{U\g-U\l} & -\Al
\end{pmatrix}.
\label{eq:transformation_matrices:dwdv}
\end{align}%
\label{eq:transformation_matrices}%
\end{subequations}%

\end{document}